\documentclass{article}

\usepackage{epsfig}
\usepackage{amsfonts}
\usepackage{geometry}

\def\a{\alpha}
\def\ve{\varepsilon}
\def\G{\Gamma}
\def\rar{\rightarrow}
\def\le{\left(}
\def\ri{\right)}
\def\no{\nonumber}
\def\pd{\partial}

\begin{document}

\begin{titlepage}
\flushright{DESY-12-087}
\vskip 2cm
\begin{center}
{\Large \bf  Solution to Bethe-Salpeter equation \\
\vskip 2mm  
via Mellin-Barnes transform}\\
\vskip 1cm  
Pedro Allendes $^{(a)},$ Bernd Kniehl $^{(b)},$  Igor Kondrashuk $^{(c)},$   \\ 
\vskip 2mm
Eduardo A. Notte Cuello $^{(d)},$    Marko Rojas Medar $^{(c)}$ \\
\vskip 5mm  
{\it  (a) Departamento de F\'\i sica, Universidad de Concepci\'on, Casilla 160-C, Concepci\'on, Chile} \\
{\it  (b) II. Institut fur Theoretische Physik, Universitat Hamburg, \\
          Luruper Chaussee 149, 22761 Hamburg, Germany} \\
{\it  (c) Departamento de Ciencias B\'asicas,  Universidad del B\'\i o-B\'\i o, \\ 
          Campus Fernando May, Casilla 447, Chill\'an, Chile} \\
{\it  (d) Departamento de Matem\'aticas, Facultad de Ciencias, Universidad de La Serena, \\ 
          Av. Cisternas 1200, La Serena, Chile}
\end{center}
\vskip 10mm

\begin{abstract}
We consider Mellin-Barnes transform of triangle ladder-like scalar diagram in $d=4$ dimensions. It is shown how    
the multi-fold MB transform of the momentum integral corresponding to an arbitrary number of rungs 
is reduced to the two-fold MB transform. For this purpose we use Belokurov-Usyukina 
reduction method for four-dimensional scalar integrals in the position space.
The result is represented in terms of Euler $\psi$-function and its derivatives.
We derive new formulas for the MB two-fold integration in  complex planes of 
two complex variables. We demonstrate that these formulas solve Bethe-Salpeter equation. 
We comment on further applications of the solution to the Bethe-Salpeter equation for the vertices 
in ${\cal N}=4$ supersymmetric Yang-Mills  theory.   
We show that the recursive property of the MB transforms observed in the present work 
for that kind of diagrams has nothing to do with quantum field theory, theory of integral transforms, 
or with theory of polylogarithms in general, but has an origin in a simple recursive property 
for smooth functions which can be shown by using basic methods of mathematical analysis.  
\end{abstract}
\end{titlepage}

\section{Introduction}

In this paper we study Mellin-Barnes (MB) transform of three-point scalar ladder-like integrals in $d=4$ space-time 
dimensions. These integrals contribute to Green's functions of four-dimensional massless scalar theories 
and can be represented in terms of UD functions \cite{Usyukina:1992jd,Usyukina:1993ch}. As has been proven in Refs.  
\cite{Kondrashuk:2008xq,Kondrashuk:2008ec,Kondrashuk:2009us} at the level of graphs, the UD functions 
transform to themselves under Fourier transform. Also, this property can be proven by making a use 
of the MB transform \cite{Allendes:2009bd}.

As has been shown in Ref. \cite{Kondrashuk:2009us},   any other scalar three-point Green's function 
even in the non-integer number  $d=4-2\ve$ of dimensions possesses such a property of invariance 
with respect to Fourier transform too. This happens due to the fact that any three-point function 
can diagrammatically be considered 
as a net of consequent three-point or four-point integrations in the position space since any 
of those integrations in the position space can be represented in terms of MB integral transforms 
with powers of spacetime intervals in denominators depending on the MB transform parameters.  
Doing consequently momentum integration via the MB transform, we come to an expression in which 
there is no momentum integral but there are integrations in complex planes of the MB transform
parameters. We can apply arguments of Ref. \cite{Kondrashuk:2009us,Allendes:2009bd} to the expressions of such a type
in order to demonstrate their invariance with respect to the Fourier transformation.

In general, technique of the MB transform is a powerful method to perform complicate multiloop calculations 
in quantum field theory \cite{Smirnov,Boos:1990rg,Bern:2005iz,DelDuca:2009au}. 
For example, the three-point one-loop function 
in the position space in the massless scalar theory is a combination of Appell's hypergeometric functions 
\cite{Davydychev:1992xr}, which have appeared as a result of residue calculation via the MB transform.
In Ref. \cite{Allendes:2009bd} it has been shown that for the three-point integrations in the position space  
with the powers of space-time intervals in the denominators equal to the integer numbers shifted by some multipliers of $\ve,$ 
where $\ve$ is a parameter of dimensional regularization, $d=4-2\ve,$ these Appell's hypergeometric functions can be 
reduced to a combination of the UD functions. On the other side, the four-point function in the position space 
in the scalar theory in $d=4$ dimensions  can be represented in terms of the UD functions too \cite{doklad}.  
The invariance of those four-point ladder integrals under the Fourier transform can be proved by means of the MB transform too 
\cite{doklad}.  
All these properties make UD functions very attractive for further investigation by making a use of the MB technique.  
The main advantage of this technique that we use in the present paper is that the three-point $d$-dimensional 
integration is transformed to a power-like form, which can be integrated in further loops as a three-point 
integration with the powers of space-time intervals in denominators which depend on the complex variables of the MB transform.

The first aim of this paper is to investigate further the properties of the UD functions via 
the MB transform. As a by-product, we derive new formulas for the MB two-fold integration in the complex planes 
of two variables of integration. The third aim is to demonstrate  compatibility between  these formulas 
for the MB transform and the Bethe-Salpeter equation for the infinite sum of triangle-like and box-like 
scalar ladders.

The fourth motivation for the investigation developed in this paper is to apply the experience obtained 
with studying the Bethe-Salpeter equation for the scalar ladders to calculation  of an  auxiliary 
vertex or Green function, which in  ${\cal N} = 4$ supersymmetric 
Yang-Mills theory does not depend on any scale, ultraviolet or infrared,
in the limit of removing dimensional regularization, $\ve=0,$ in $d=4.$
That vertex should satisfy the Bethe-Salpeter equation, particular for this Green's function. 
This is $Lcc$ vertex in which an auxiliary field $L$ couples to two (self-adjoint) Faddeev-Popov ghost 
fields $c.$ To investigate that vertex by the Bethe-Salpeter equation has a sense since the
ghost fields are  the only fields of this vertex which stand in the measure of the path integral, 
and therefore this equation restricts the vertex strongly.

As has been shown in Refs. \cite{Cvetic:2004kx,Kondrashuk:2004pu,Cvetic:2006kk}, 
in ${\cal N} = 4$ supersymmetric Yang-Mills theory starting with this vertex all the other correlators 
can be found in the four-dimensional limit (when $d=4$) by using Slavnov-Taylor identity 
\cite{Slavnov:1972fg,Taylor:1971ff,Slavnov:1974dg,Faddeev:1980be,Lee:1973hb,
Zinn-Justin:1974mc,Becchi:1974md,Tyutin:1975qk}. 
The explicit two-loop result for that vertex has been found in Refs.~\cite{Cvetic:2006iu,Cvetic:2007fp,Cvetic:2007ds}.
All the poles in $\ve$ disappear in all number of loops for this vertex, however 
it cannot be analysed by the methods of conformal field theory since the auxiliary field $L$ 
does not propagate. Nevertheless,  the three-point function of dressed mean gluons of Refs. 
\cite{Cvetic:2004kx,Kondrashuk:2004pu,Cvetic:2006kk} can be fixed by conformal symmetry  in analogy to 
\cite{Mitra:2008yr,Mitra:2008pw,Mitra:2009zm,Freedman:1998tz,Erdmenger:1996yc}, however, we cannot 
find higher-point correlators of the dressed mean fields from this Green's function by using ST identity.

As it can be seen from Refs. \cite{Cvetic:2004kx,Kondrashuk:2004pu,Cvetic:2006kk}, 
the explicit structure of $Lcc$ correlator includes logarithms and the UD functions of ratios of spacetime 
distances between arguments of the Green's functions.     
As it follows from  Ref.\cite{Kondrashuk:2008xq}, the ladder-like diagrams  with horizontal gluon lines,
that are important subclass of all the diagrams contributing to that vertex  can be represented 
in terms of the UD functions too for an arbitrary number of the horizontal gluon lines. In the next paper we will 
demonstrate that any of the contributions to the correlator can be represented in terms of the UD functions 
\cite{next}.

In this paper we consider a particular solution to the Bethe-Salpeter equation, that represents an infinite 
sum of the triangle scalar ladder diagrams in four space-time dimensions in the MB representation.
This particular solution has already been studied in the case of the box-ladder diagrams 
by Broadhurst and Davydychev in Ref. \cite{Broadhurst:2010ds}. The authors of \cite{Broadhurst:2010ds}
 calculated a sum of those box-ladder 
diagrams by making a use of the integral representation for the UD functions of Ref. \cite{Usyukina:1993ch}. As has been shown 
in Refs.\cite{Usyukina:1992jd,Usyukina:1993ch}, the triangle-like and the box-like ladders in four space-time 
dimensions are related by conformal transformation in dual space and their MB transforms coincide.  
The difference of the present paper with the approach of  Ref. \cite{Broadhurst:2010ds} is that we do the calculation 
via the MB transformation.  We derive recursive relations for the MB transforms of the momentum integrals corresponding 
to the UD triangle diagrams by using Belokurov-Usyukina  
trick \cite{Belokurov:1983km} for the ladder diagrams in the position space, reducing the multi-fold MB integrals 
to the two-fold MB integrals.

\section{Loop reduction in $d = 4$ dimensions}

\begin{figure}[ht!] 
\centering\includegraphics[scale=0.27]{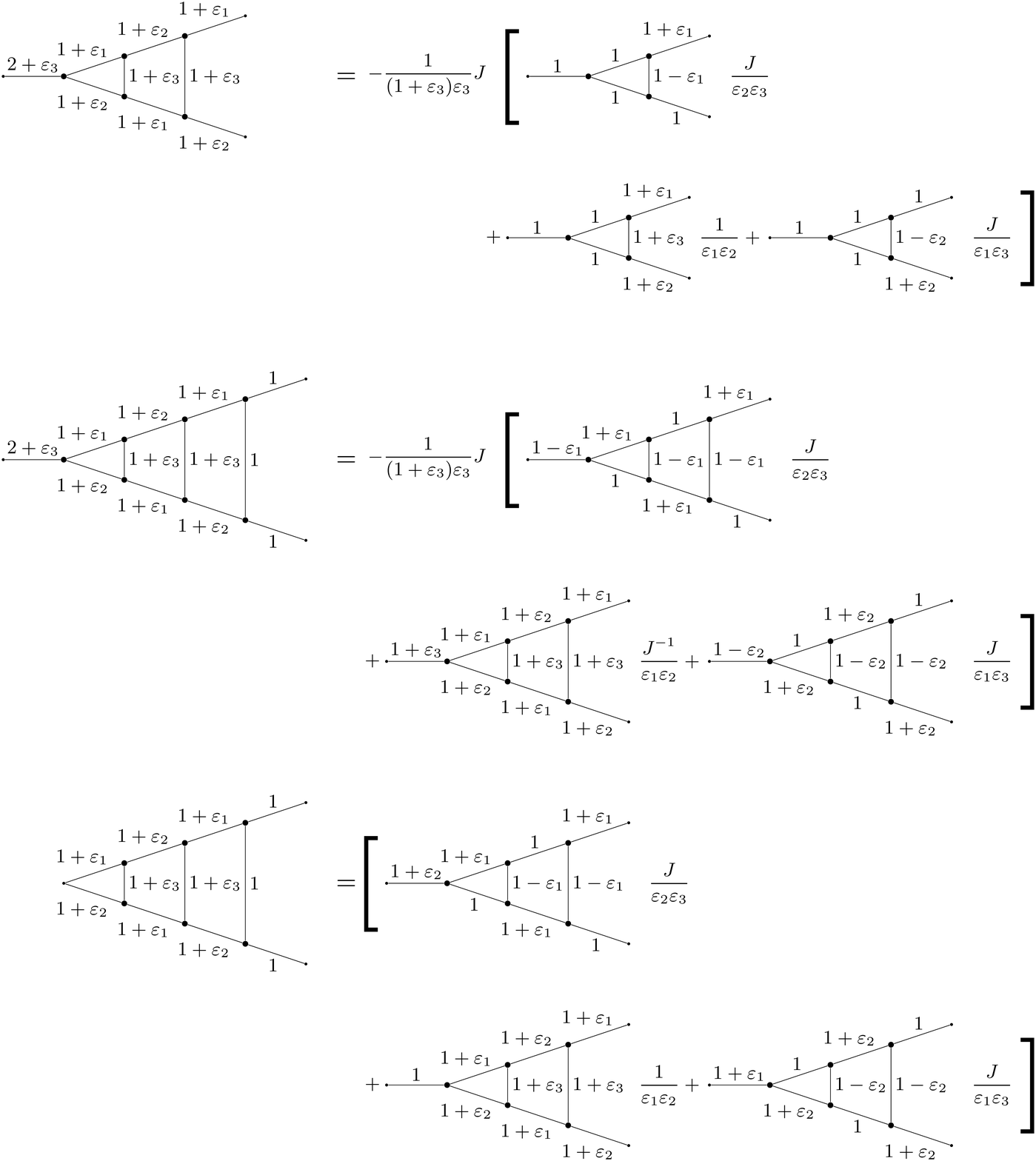}
\caption{\footnotesize  Basic relations}
\label{figure-1}
\end{figure}

\begin{figure}[ht!] 
\centering\includegraphics[scale=0.27]{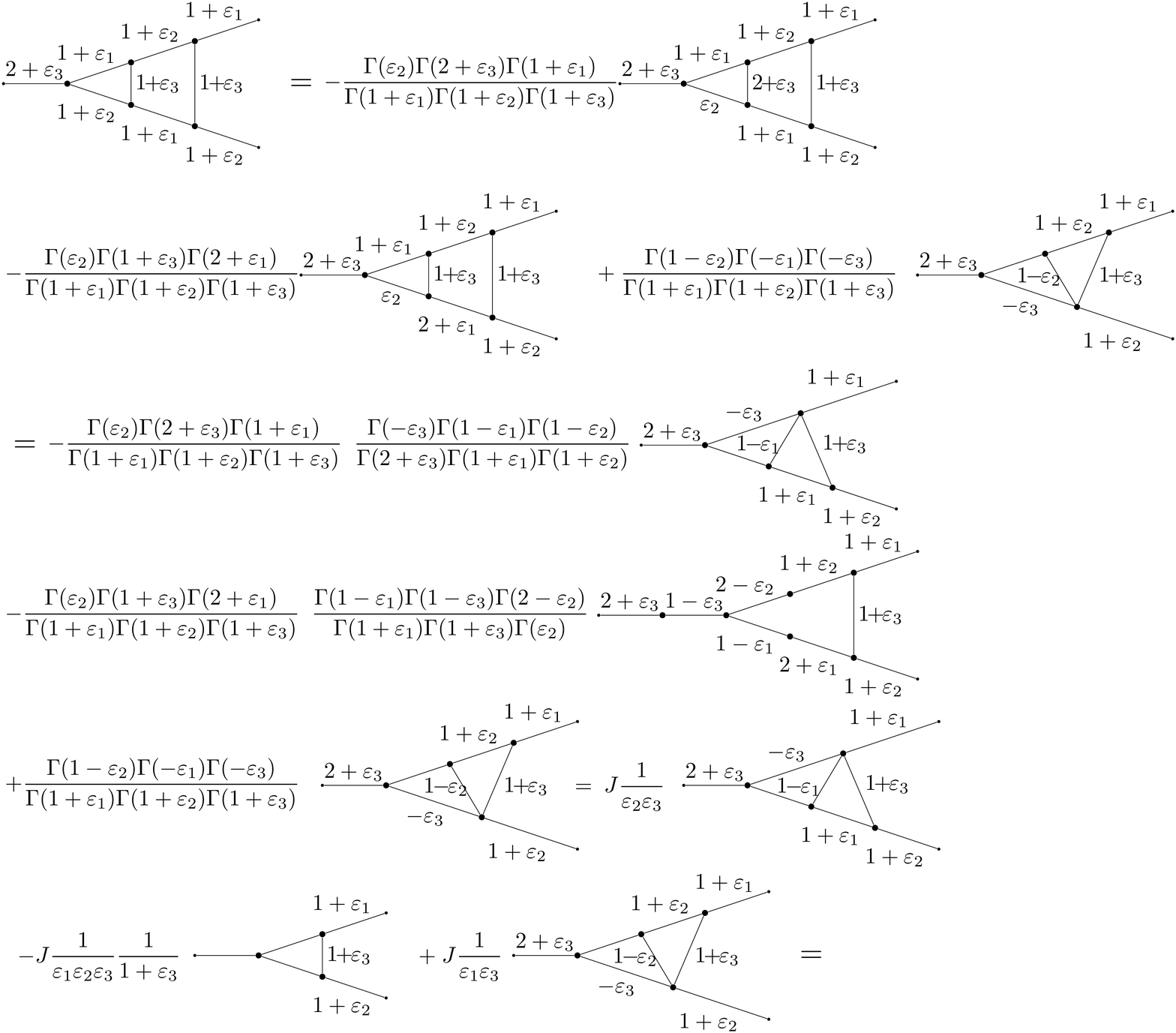}
\caption{\footnotesize  Derivation of the basic relation. Part I.
Lines without any index are the lines with index 1 }
\label{figure-2}
\end{figure}

\begin{figure}[ht!!] 
\centering\includegraphics[scale=0.21]{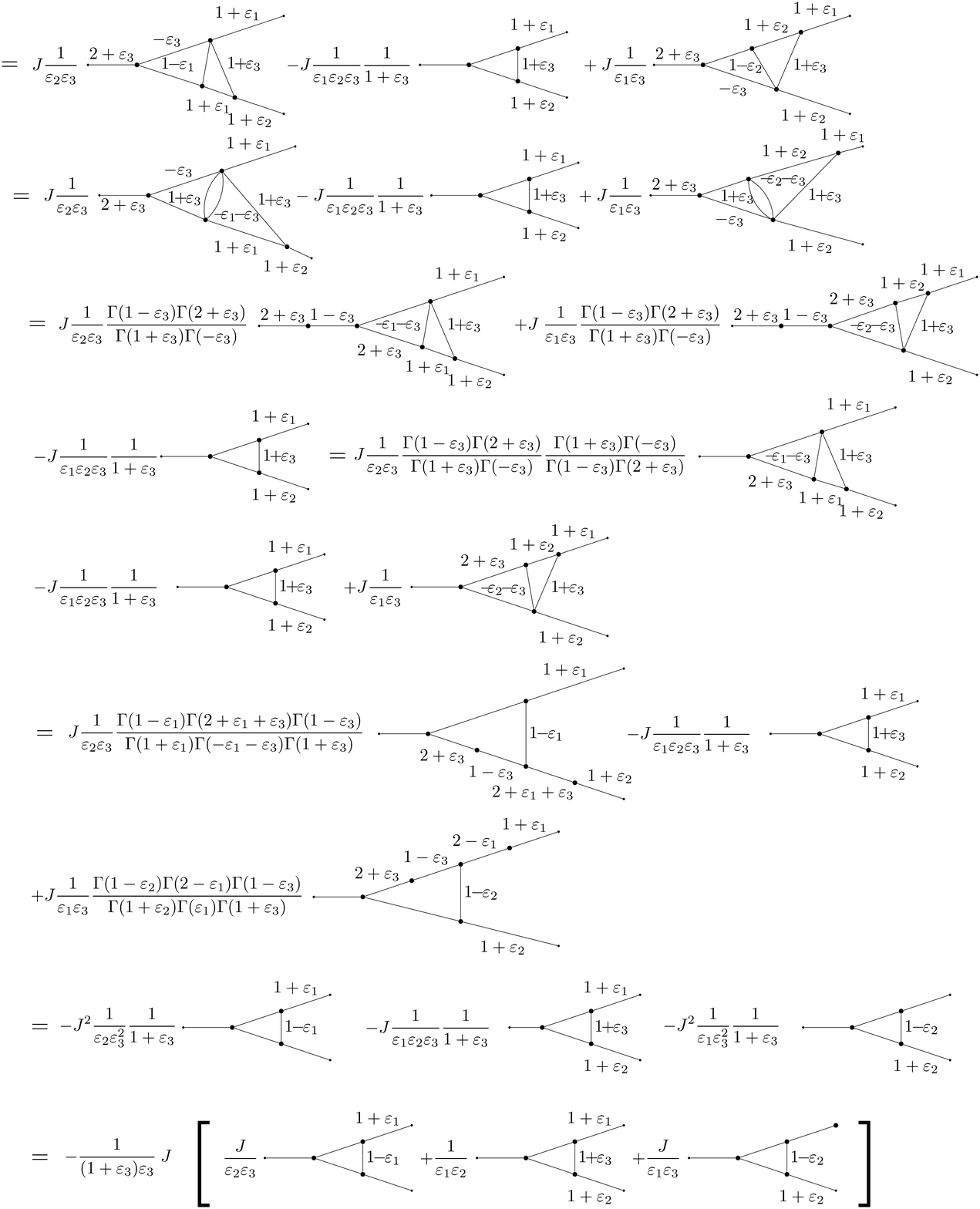}
\caption{\footnotesize  Derivation of the basic relation. Part II. Lines without any index are the lines with index 1}
\label{figure-3}
\end{figure}

\begin{figure}[ht!] 
\centering\includegraphics[scale=0.27]{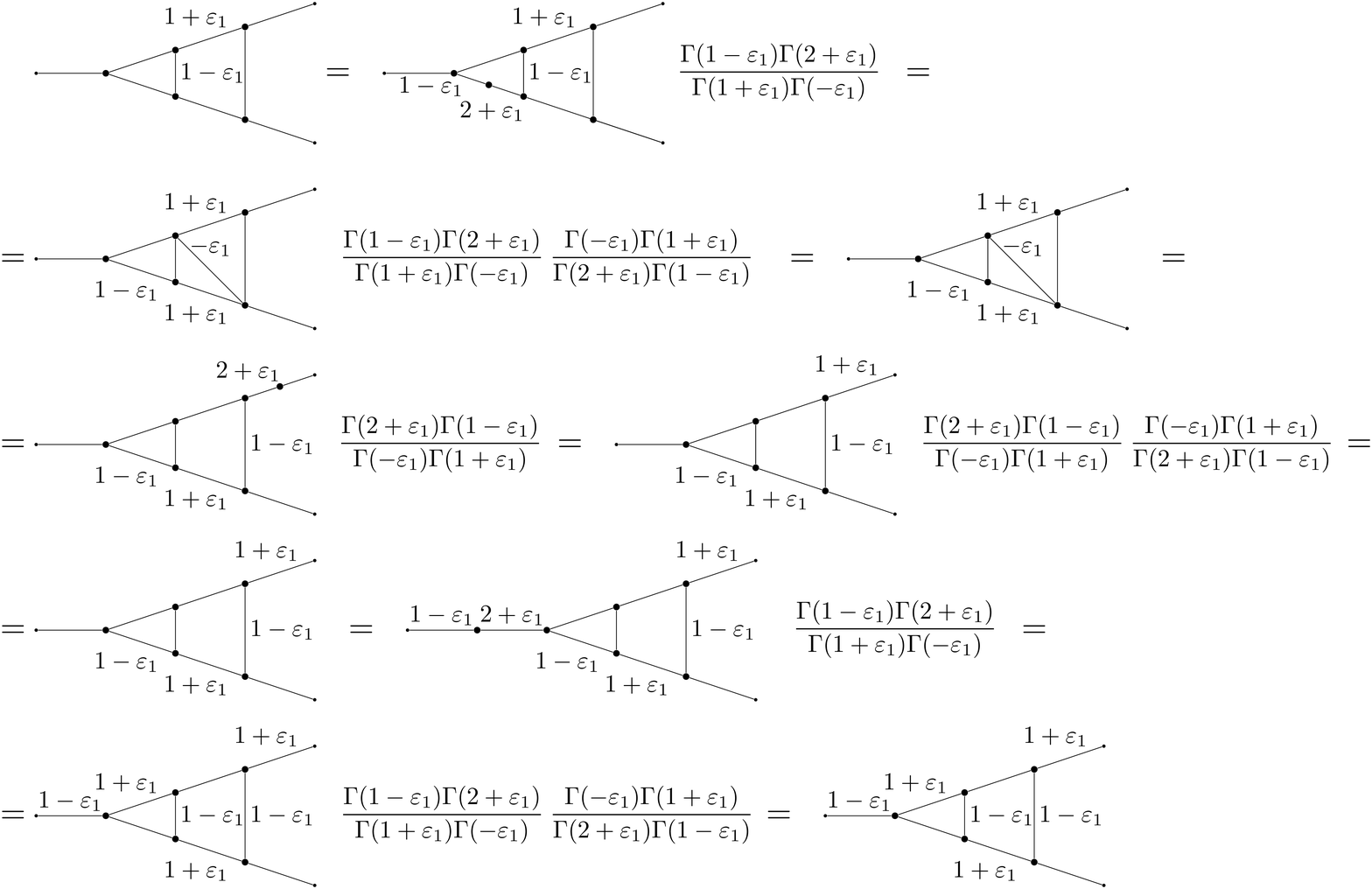}
\caption{\footnotesize  Transformation of the first diagram in Fig. (1).
Lines without any index are the lines with index 1}
\label{figure-4}
\end{figure}

\begin{figure}[ht!] 
\centering\includegraphics[scale=0.27]{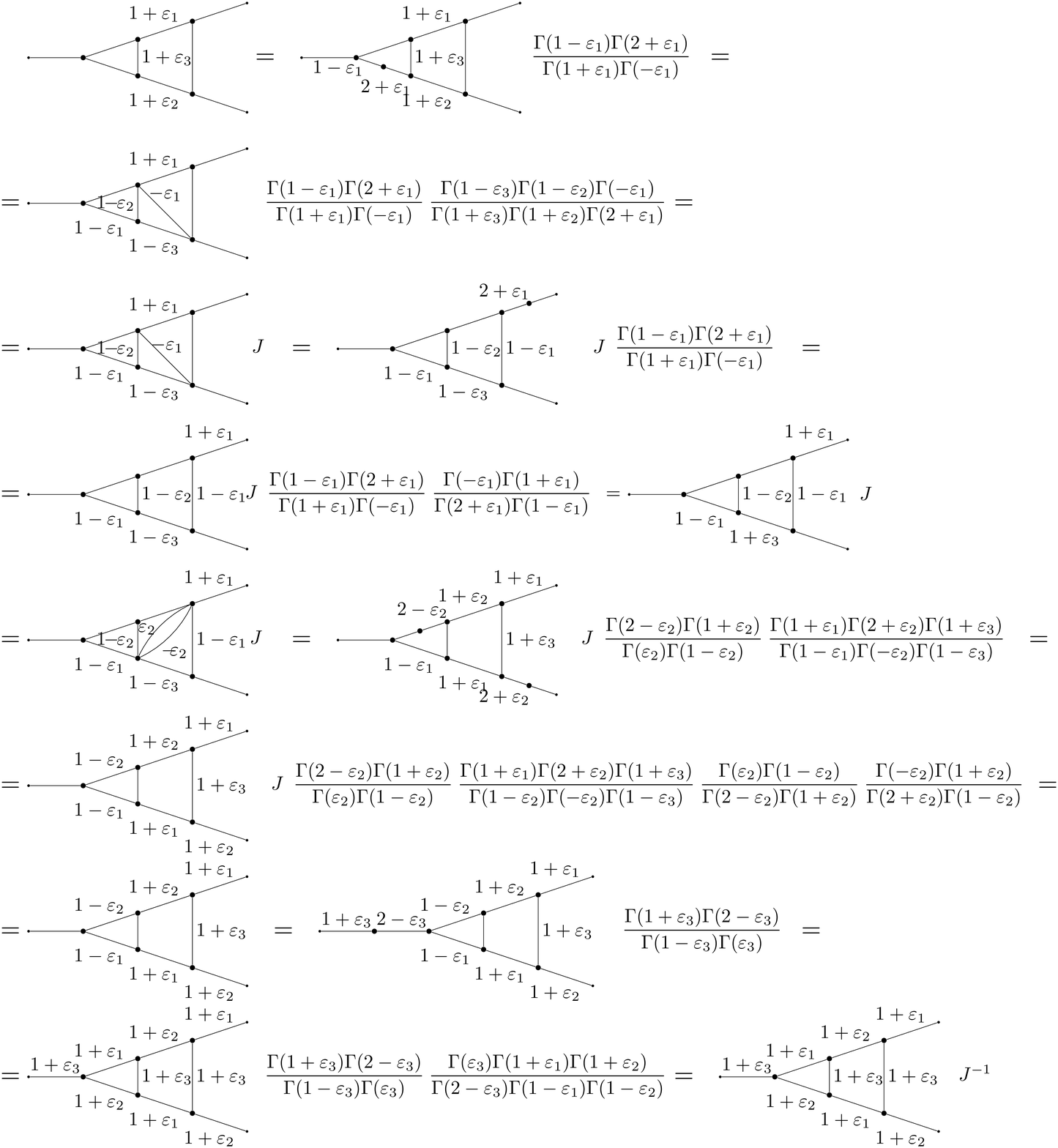}
\caption{\footnotesize  Transformation of the second diagram in Fig. (1)}
\label{figure-5}
\end{figure}

\begin{figure}[ht!] 
\centering\includegraphics[scale=0.27]{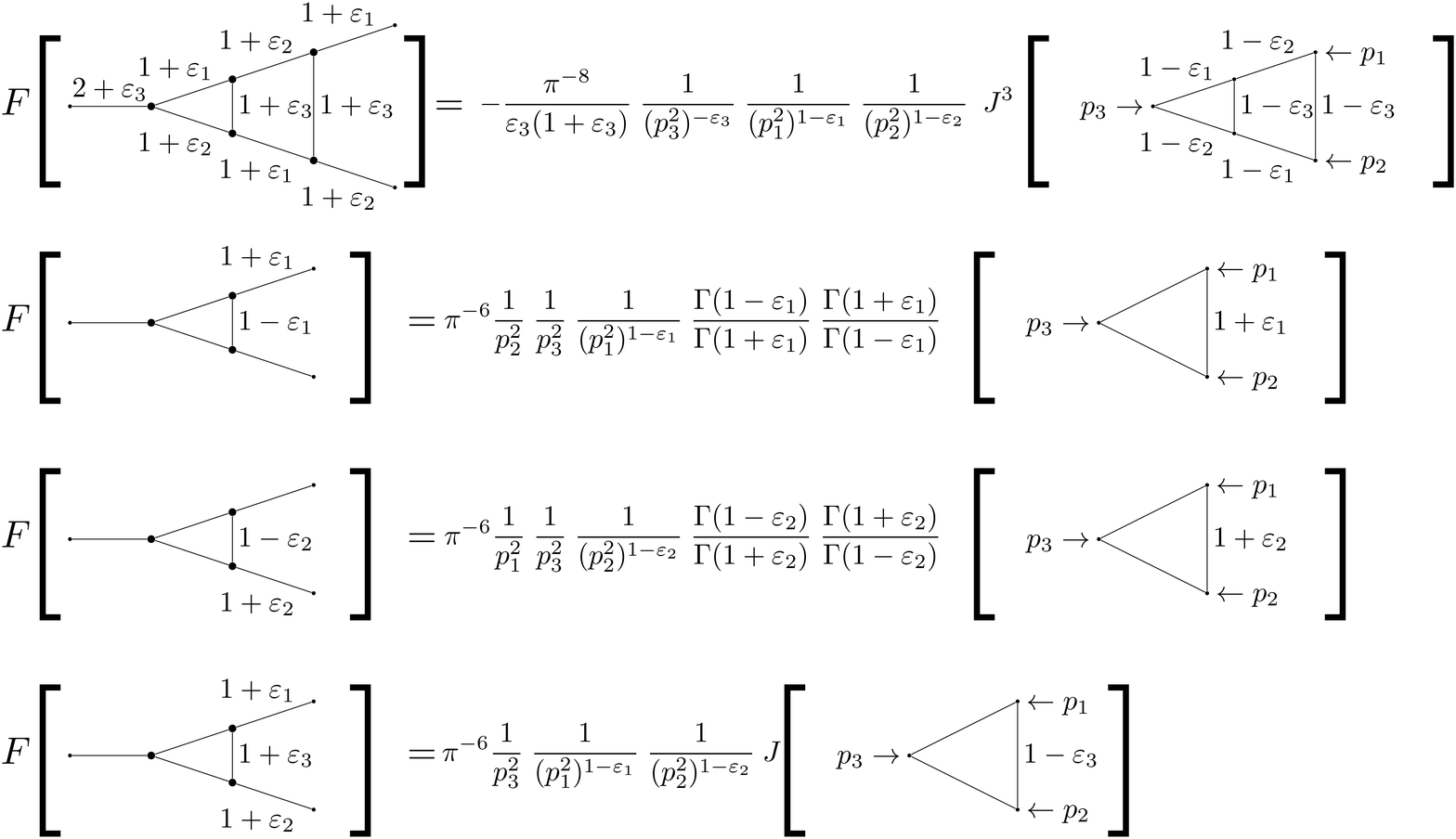}
\caption{\footnotesize Fourier transform of the diagrams in Fig. (1). The r.h.s. corresponds to the momentum integrals with indicated indices.}
\label{figure-6}
\end{figure}

\begin{figure}[ht!] 
\centering\includegraphics[scale=0.27]{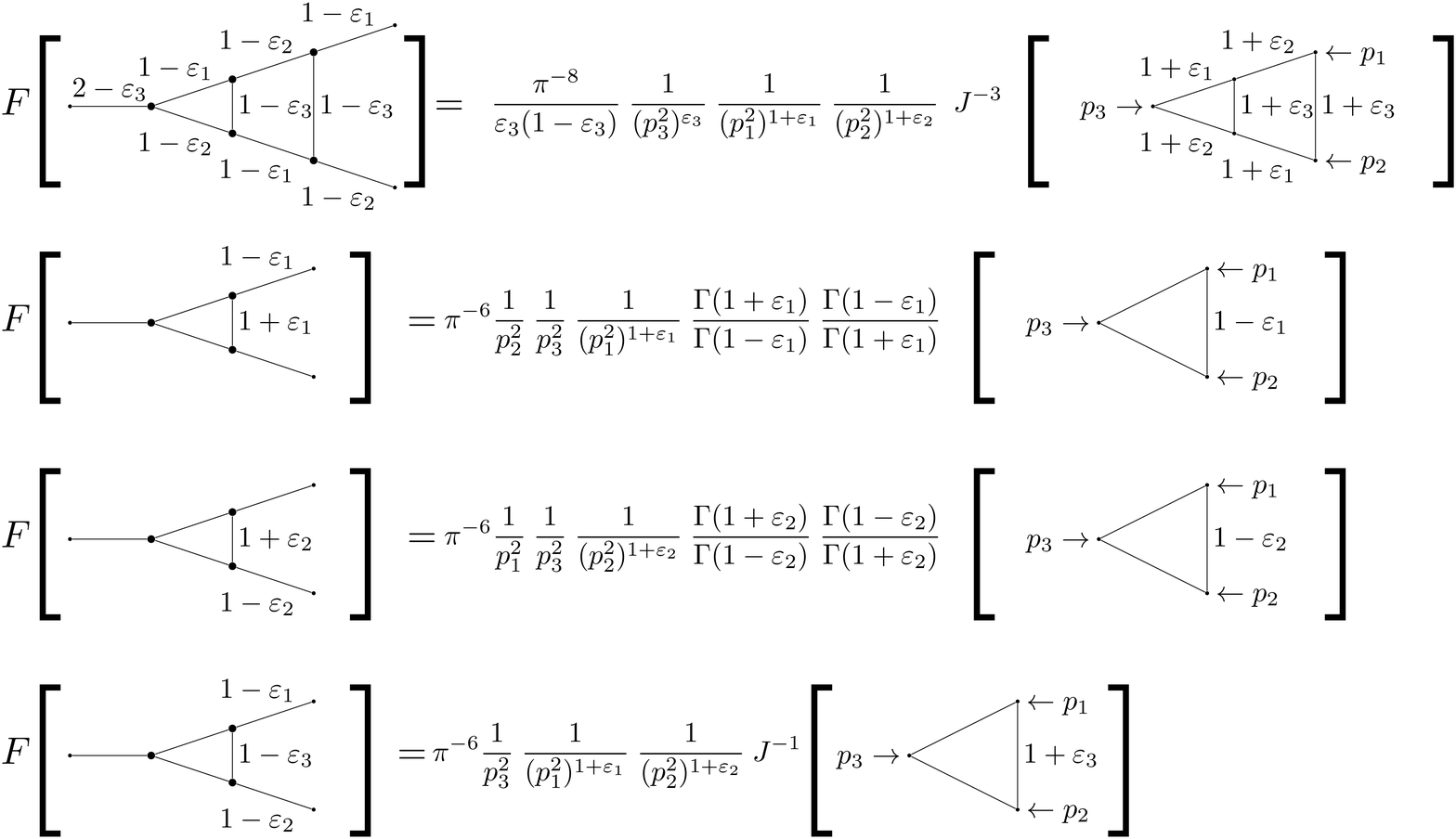}
\caption{\footnotesize Fourier transform of the diagrams in Fig. (1)}
\label{figure-7}
\end{figure}

\begin{figure}[ht!] 
\centering\includegraphics[scale=0.40]{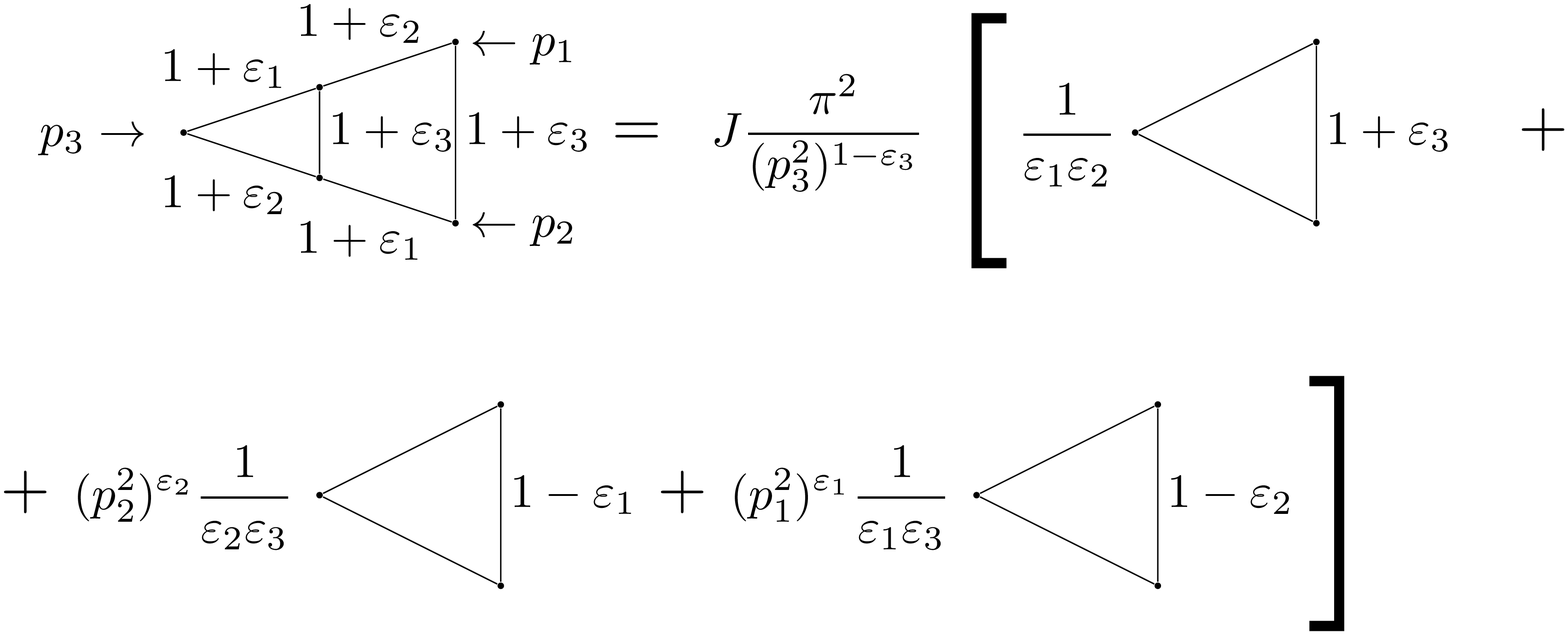}
\caption{\footnotesize  Formula (25) of Ref. \cite{Usyukina:1992jd}}
\label{figure-8}
\end{figure}

In this section we describe a trick with a help of which the triangle ladder diagram of $n$ loops is reduced 
to the triangle ladder diagram of $n-1$ loops. This trick has been published for the first time in  paper 
\cite{Belokurov:1983km}. The  effect of the loop reduction has earlier been discovered in 
Ref. \cite{Usyukina:1983gj}
for a propagator-type diagram of  a particular topology in a special limit for the indices of diagram. 
Note, that in Ref. \cite{Belokurov:1983km} the derivation has been performed without that special limit.
Later, the key intermediate point of the trick has been published in Ref. \cite{Usyukina:1991cp}. 
Since all of that together has never been published in one article in detail, we  
collect all the results and intermediate steps we need for deriving our formulas for the MB transform
of the triangle-ladder diagram.  

The  basic formulas  we used to derive these diagrammatic relations are the chain integration for an any $\a_1$ and $\a_2,$ except for 
$d/2 +n,$ $n \in N,$
\begin{eqnarray}
\int Dx \frac{1}{\left[(x_1-x)^2\right]^{\a_1} \left[(x_2-x)^2\right]^{\a_2}} = 
\frac{A(\a_1,\a_2,d-\a_1-\a_2)}{\left[(x_1-x_2)^2\right]^{\a_1 + \a_2 - d/2}}, \label{remove-base}
\end{eqnarray}
where 
\begin{eqnarray*}
A(\a_1,\a_2,\a_3)= \frac{\G(d/2-\a_1)\G(d/2-\a_2)\G(d/2-\a_3)}{ \G(\a_1) \G(\a_2) \G(\a_3)},
\end{eqnarray*}
and the star-triangle relation for $\a_1 + \a_2 + \a_3 = d$
\begin{eqnarray*}
\int Dx \frac{1}{\left[(x_1-x)^2\right]^{\a_1} \left[(x_2-x)^2\right]^{\a_2} \left[(x_3-x)^2\right]^{\a_3}   } = \\
\frac{A(\a_1,\a_2,\a_3)}{\left[(x_1-x_2)^2\right]^{d/2-\a_3}\left[(x_2-x_3)^2\right]^{d/2-\a_1} \left[(x_1-x_3)^2\right]^{d/2-\a_2}}.
\end{eqnarray*}
A new $d$-dimensional measure, introduced in ref. \cite{Cvetic:2006iu},
\begin{eqnarray*}
Dx \equiv \pi^{-\frac{d}{2}}d^d x 
\end{eqnarray*}
is used to avoid powers of $\pi$ on the r.h.s. of the diagrams.

The bold dots correspond to the internal vertices of the diagrams, and this measure corresponds to the bold dots.
We make an integration over coordinates of the internal vertices. 
In the first line of fig. (\ref{figure-1}) we present the formula which we want to derive.
This formula is present in paper \cite{Usyukina:1991cp}, however with a bit different 
indices of the diagrams. Later we will see how the r.h.s. of the first line of fig. (\ref{figure-1}) can be put 
in equivalence to fig. (4) of  \cite{Usyukina:1991cp}. The second and the third lines of fig. (\ref{figure-1})
are consequences of the first line. How exactly the second and the third lines can be derived 
from the first line we explain further in the present paper. Usually, it can be done by inserting points of integration into the lines 
of the graphs. The third line is published in Ref. \cite{Belokurov:1983km}.
The second line has never been published.  The condition for the $\ve$-terms in the indices is 
\begin{eqnarray*}
\ve_1 + \ve_2 + \ve_3 = 0. 
\end{eqnarray*}

In fig.(\ref{figure-2}) and in fig.(\ref{figure-3}), the latter is a continuation of fig.(\ref{figure-2}), 
we have shown how the first line of fig.(\ref{figure-1}) can be derived. Actually, the only 
mathematics that has been used here is the uniqueness star-triangle relation and the integration by parts,
both the relations are well-known in the scalar field theories \cite{Unique,Vasiliev:1981dg,Vasil} 
(for a short review, Ref. \cite{Kazakov:1984bw}). In all the figures from  (\ref{figure-1}) till 
(\ref{figure-9}) we use for factor $J$ the definition 
\begin{eqnarray*}
J =   \frac{\G(1-\ve_1)\G(1-\ve_2) \G(1-\ve_3) }{\G(1+\ve_1)\G(1+\ve_2) \G(1+\ve_3) }, 
\end{eqnarray*}
taken from Ref. \cite{Belokurov:1983km}.

In fig. (\ref{figure-4}) we show how the first diagram on the r.h.s. of the first line 
of fig. (\ref{figure-1}) transforms to the first diagram on the r.h.s. of the second line 
of figure  (\ref{figure-1}). Again, we converted  non-unique 
vertices to unique vertices by inserting the point of integration into one of the propagators attached to the vertex. 
Then, the star-triangle relation has been applied in a direct way or in a triangle-star way, until 
the diagram is converted 
to a diagram with desirable values of indices. The line without any index means that the index of the line 
is 1. We put the exact value of this index on the corresponding propagators of fig. (\ref{figure-1}) only. 
In the rest of figures we omit the index 1 just not to overload the diagram with indices.

In fig. (\ref{figure-5}) the second diagram of the first line of fig. (\ref{figure-1}) 
is transformed to the second diagram of the second line of fig. (\ref{figure-1}). Again, nothing else but 
the creating of unique vertices and triangles by putting new points of integration into the propagators has been 
used to obtain the desired indices. Thus, we have proven the first line and the second line of fig. (\ref{figure-1}).

The third line of fig. (\ref{figure-1}) can be obtained from the second line of fig. (\ref{figure-1})
by simple integration. Indeed, we  eliminate the leftmost propagator on the l.h.s. of the second line 
by convoluting its leftmost point with the line whose index is $2-\ve_3,$
\begin{eqnarray}
\int~d^4x \frac{1}{\left[(x_1-x)^2\right]^{2-\ve_3} \left[(x_2-x)^2\right]^{2+\ve_3} } = \pi^d
\frac{\G(\ve_3)}{\G(2-\ve_3)}\frac{\G(-\ve_3)}{\G(2+\ve_3)}\delta^{(4)}(x_1-x_2). \label{remove}
\end{eqnarray}

To derive formula (\ref{remove}) we need  to replace in Eq. (\ref{remove-base})  each factor in the integrand with its 
integral Fourier transform. The formulas for the Fourier transform of the factor like those can be found 
in Ref. \cite{Kazakov:1984bw}. Indeed, since
\begin{eqnarray*}
\int~d^d p ~e^{ip x} \frac{1}{(p^2)^\a} = \pi^{d/2}\frac{\G(d/2-\a)}{\G(\a)}\left(\frac{4}{x^2}\right)^{d/2-\a}  \Longrightarrow  
\end{eqnarray*}
\begin{eqnarray*}
\frac{1}{(x^2)^{\a}}  = \pi^{-d/2} 4^{-\a}  \frac{\G(d/2-\a)}{\G(\a)}\int~d^d p ~e^{ip x} \frac{1}{(p^2)^{d/2-\a}}, 
\end{eqnarray*}
we can write under the condition $\a_1 + \a_2 = d$ 
\begin{eqnarray*}
\int~d^dx \frac{1}{\left[(x_1-x)^2\right]^{\a_1} \left[(x_2-x)^2\right]^{\a_2}} = \\
\pi^{-d}4^{-\a_1-\a_2}\frac{\G(d/2-\a_1)}{\G(\a_1)}\frac{\G(d/2-\a_2)}{\G(\a_2)} \int~d^d x~d^d p~d^d q \frac{e^{ip (x-x_1)} e^{iq (x-x_2)}}{ (p^2)^{d/2-\a_1}(q^2)^{d/2-\a_2} } = \\
\pi^{-d}4^{-\a_1-\a_2}\frac{\G(d/2-\a_1)}{\G(\a_1)}\frac{\G(d/2-\a_2)}{\G(\a_2)} (2\pi)^d\int~d^d p \frac{e^{ip (x_1-x_2)}}{ (p^2)^{d-\a_1-\a_2}} = \\
\pi^{-d}4^{-\a_1-\a_2}\frac{\G(d/2-\a_1)}{\G(\a_1)}\frac{\G(d/2-\a_2)}{\G(\a_2)} (2\pi)^{2d}\delta^{(d)}(x_1-x_2) = \\
\pi^{d}\frac{\G(d/2-\a_1)}{\G(\a_1)}\frac{\G(d/2-\a_2)}{\G(\a_2)} \delta^{(d)}(x_1-x_2).
\end{eqnarray*}
As a consequence, we obtain in case of $\a_1 + \a_2 = d$ 
\begin{eqnarray*}
\int~Dx \frac{1}{\left[(x_1-x)^2\right]^{\a_1} \left[(x_2-x)^2\right]^{\a_2}} = \pi^{d/2}\frac{\G(d/2-\a_1)}{\G(\a_1)}\frac{\G(d/2-\a_2)}{\G(\a_2)} \delta^{(d)}(x_1-x_2).
\end{eqnarray*}

\subsection{Description of Fourier Transform for the diagrams}

In fig. (\ref{figure-6}) the Fourier transform is made for all the diagrams that appear in the first line 
of fig. (\ref{figure-1}). The procedure of Fourier transformation is done as follows. We replace each factor in 
the integrand of the position space representation with the integral Fourier transform of the corresponding factor 
in the momentum space representation. Space-time coordinates appear in the exponentials of the Fourier transforms. 
Integrating over the coordinates of internal vertices we create Dirac $\delta$-functions, corresponding to the 
momentum conservation in each vertex of integration in the position space (internal vertex). The  
momentum integrals over loop momenta will be the Fourier transforms of the integrals in the position space. 
For example, the described procedure in a particular case of the one-rung integral in the position space would 
result in  
\begin{eqnarray*}
\frac{1}{[31]}\int~d^4y~d^4z \frac{1}{[2y][1y][3z][yz][2z]}= 
\frac{1}{(2\pi)^4}\int~d^4p_1d^4p_2d^4p_3 ~ \delta(p_1 + p_2 + p_3) \times\no\\
\times e^{ip_2x_2} e^{ip_1x_1} e^{ip_3x_3} C^{(2)}(p_1^2,p_2^2,p_3^2). 
\end{eqnarray*} 
We assume the notation of Ref. \cite{Cvetic:2006iu},  where $[Ny]= (x_N - y)^2$ and analogously for $[Nz]$ and $[yz],$ 
that is, $N=1,2,3$ stands for $x_N=x_1,x_2,x_3,$ respectively, which are the external points of the triangle-like 
ladder diagram.   In Refs. \cite{Usyukina:1992jd,Usyukina:1993ch} $C^{(n)}(p_1^2,p_2^2,p_3^2)$ is the definition for
the result of momentum integrals for $n$-rung triangle ladder diagram in the momentum space representation.
Just to make clear the definition for the operation of Fourier transformation   used in fig.  (\ref{figure-6}),
we provide an example of the relation  
\begin{eqnarray*}
F\left[\frac{1}{[31]}\int~d^4y~d^4z \frac{1}{[2y][1y][3z][yz][2z]} \right] = \frac{1}{(2\pi)^4}C^{(2)}(p_1^2,p_2^2,p_3^2).
\end{eqnarray*}
The transformation from the position space to the momentum space is necessary to normalize on the results for 
the MB transform of the UD functions done in \cite{Usyukina:1992jd,Usyukina:1993ch} for the momentum space integrals.

Fig. (\ref{figure-7}) is fig. (\ref{figure-6}) with signs changed for all $\ve$-terms. After this change of signs,
the Fourier transform of the first line of fig. (\ref{figure-1}) takes the form depicted in fig. (\ref{figure-8}).
In that diagrammatic relation we recognize Eq. (25) of Ref. \cite{Usyukina:1992jd}.

\section{Recursive formula for MB transform of the UD functions}

\begin{figure}[ht!] 
\centering\includegraphics[scale=0.40]{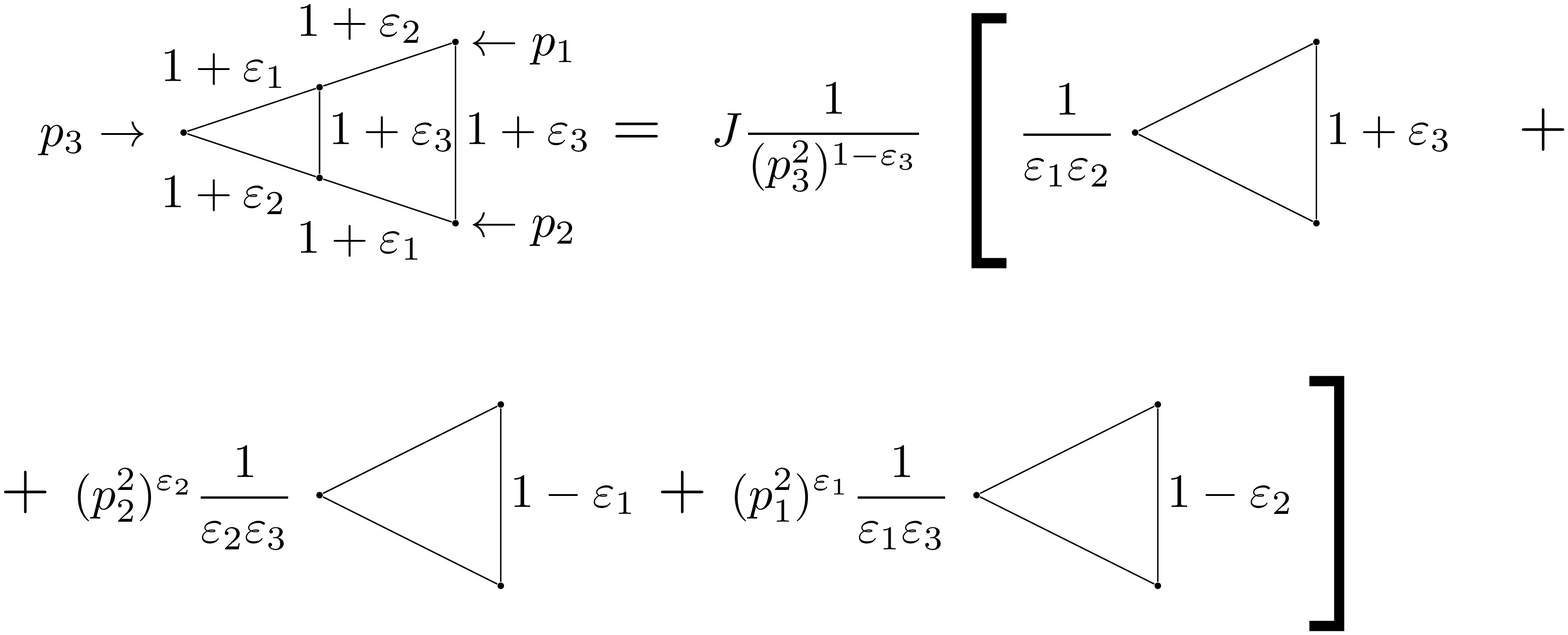}
\caption{\footnotesize  Fig. (\ref{figure-8}) after measure redefinition  (\ref{k-measure})}
\label{figure-9}
\end{figure}

We start this section with writing the definition of the momentum integral for the diagram in the r.h.s. of Fig. (\ref{figure-8}). 
To define the integral measure in the momentum space we use the notation 
\begin{eqnarray}
Dk \equiv \pi^{-\frac{d}{2}}d^d k.    \label{k-measure}
\end{eqnarray}
This is done to avoid powers of $\pi$ in formulas for the momentum integrals. For example, fig. (\ref{figure-8}) takes the form shown in fig. (\ref{figure-9}).

We take a formula from paper \cite{Usyukina:1992jd} for the representation of the three-point momentum integral 
\begin{eqnarray*}
J(\nu_1,\nu_2,\nu_3) = \int~Dk~\frac{1}{\left[(k + q_1)^2\right]^{\nu_1} \left[(k + q_2)^2 \right]^{\nu_2}
\left[(k + q_3)^2\right]^{\nu_3}}. 
\end{eqnarray*}
in terms of the MB transform 
\begin{eqnarray}
J(\nu_1,\nu_2,\nu_3) = \frac{1}{\Pi_{i} \G(\nu_i) \G(d-\Sigma_i \nu_i)} \frac{1}{{(p^2_3)}^{ \Sigma \nu_i -d/2}}
\oint_C dz_2~dz_3 x^{z_2}  y^{z_3}
\left\{ \G \le -z_2 \ri \G \le -z_3 \ri \right.\no\\
\left. \G \le -z_2 -\nu_2-\nu_3 + d/2 \ri \G \le -z_3-\nu_1-\nu_3 + d/2 \ri 
\G \le z_2 + z_3  + \nu_3 \ri  \G \le  \Sigma \nu_i - d/2 + z_3 + z_2 \ri \right\} \equiv \label{J-arb}\\
\equiv \frac{1}{{(p^2_3)}^{ \Sigma \nu_i -d/2}}\oint_C dz_2~dz_3 x^{z_2}  y^{z_3} 
D^{(z_2,z_3)} [\nu_1,\nu_2,\nu_3], \no
\end{eqnarray}
where we have introduced the definition of function $D,$
\begin{eqnarray}
D^{(z_2,z_3)}[\nu_1,\nu_2,\nu_3] = \frac{ \G \le -z_2 \ri \G \le -z_3 \ri \G \le -z_2 -\nu_2-\nu_3 + d/2 \ri 
\G \le -z_3-\nu_1-\nu_3 + d/2 \ri }
{\Pi_{i} \G(\nu_i) } \no\\
\times 
\frac{ \G \le z_2 + z_3  + \nu_3 \ri  \G \le  \Sigma \nu_i - d/2 + z_3 + z_2 \ri }
{\G(d-\Sigma_i \nu_i)}. \label{J-arb-2}
\end{eqnarray}

Here and in the rest of the paper we use the notation 
\begin{eqnarray*}
x \equiv \frac{p^2_1}{p^2_3},~~~ y \equiv \frac{p^2_2}{p^2_3},
\end{eqnarray*}
where the $d$-dimensional momenta $p_1,p_2,p_3$ satisfy the conservation law $p_1 + p_2 + p_3 = 0$ and 
are related to $d$-dimensional momenta $q_1,q_2,q_3$ by a parametrization  
\begin{eqnarray*}
p_1 = q_3 - q_2,\no\\ 
p_2 = q_1 - q_3,\no\\
p_3 = q_2 - q_1. 
\end{eqnarray*}
As a consequence of this definition, $p_1$ appears to be a momentum that enters the one-loop triangle 
diagram in the vertex of the triangle which is opposite to the line with index $\nu_1.$

\subsection{Description of notation in the MB representation}

We absorb into the definition of the MB transform $D^{(z_2,z_3)}[\nu_1,\nu_2,\nu_3]$ of this three-point integral
all the factors except for a power of the square of the external momentum  $p^2_3.$ For $d=4$ in the denominator 
we have a sum of the indices minus two for the power of  $p^2_3.$

We do not write the powers of $i,$ supposing that we work in Euclidean space and the corresponding 
power of $i$ can be recovered back after Wick rotation.

The contour of integration $C$ passes a bit on the left of the imaginary axis, separates left and right 
poles and should be closed to the left infinity or to the right infinity. We choose to close the  
contour of integration in the complex plane to the right infinity.  It could be closed to the left 
infinity too but it makes more complicate to take the residues into account since the residues
in variables $z_2$ and $z_3$ are mixed in that case. Whether we have to close the contour to the right infinity or to 
the left infinity, the result should be the same function. 
We omit the factor $1/2\pi i$ that accompanies each integration over  MB transform parameter. 
The inverse factor is generated in front of the residues.

It is known that this representation of the three-point integral can be derived by applying the two-fold MB transform 
to the integral over Feynman parameters, with producing the Euler beta functions \cite{Smirnov}
after integrating over these parameters. 
There is a difference with the representation used in Ref.\cite{Allendes:2009bd}. The form of 
Ref.\cite{Allendes:2009bd} can be recovered from (\ref{J-arb}) by a cyclic redefinition. 
In the rest of the paper, we use representation (\ref{J-arb}).

\subsection{Multi-fold MB transforms of the UD functions}

To develop the recursive equations for the MB transforms of the UD functions, we recover the 
definition of the MB transforms for the  triangle ladder diagrams in $d=4$ spacetime dimensions
\begin{eqnarray*}
C^{(n)}(p_1^2,p_2^2,p_3^2) =   \frac{1}{(p_3^2)^n} \Phi^{(n)}\le x,y\ri 
= \frac{1}{(p_3^2)^n} \oint_C dz_2~dz_3 x^{z_2}  y^{z_3} {\cal{M}}^{(n)} (z_2,z_3),
\end{eqnarray*}
where we used the notation   ${\cal{M}}^{(n)} (z_2,z_3)$ for the MB transform of the UD function $\Phi^{(n)}\le x,y\ri$
of two arguments. The definition of incoming momenta for the 
triangle ladder diagrams is like in Ref.\cite{Usyukina:1992jd,Usyukina:1993ch,Kondrashuk:2009us}. 
We include in the definition of ${\cal{M}}^{(n)} (z_2,z_3)$ all the factors that can appear in front of  
the MB transformation except for the power of the momentum $p_3^2.$

The two-fold MB transform is known for the first UD function only. For a UD function of a higher order, 
the result is given in terms of the multi-fold MB transform \cite{Usyukina:1992jd,Usyukina:1993ch}. 
In the next section we reduce the multi-fold MB transform to the two-fold MB transform by making a use of the loop 
reduction trick described in the previous section. 

The iterative integral relation for the triangle ladder diagram given in Refs. 
\cite{Usyukina:1992jd,Usyukina:1993ch,Kondrashuk:2009us},
\begin{eqnarray*}
C^{(n)}(p_1^2,p_2^2,p_3^2) =  \int~d^4r_n~\frac{C^{(n-1)}((p_1 + r_n)^2,(p_2 - r_n)^2, p_3^2 )}
{(p_1 + r_n)^2 (p_2 - r_n)^2 r_n^2}, 
\end{eqnarray*}
results in an integral recursive relation for the MB transforms of the UD functions 
\begin{eqnarray*}
\Phi^{(n+1)}\le \frac{p_1^2}{p_3^2},\frac{p_2^2}{p_3^2}\ri =  
p_3^2\int~d^4r~\Phi^{(n)}\le \frac{(p_1+r)^2}{p_3^2},\frac{(p_2-r)^2}{p_3^2}\ri
\frac{1}{(p_1 + r)^2 (p_2 - r)^2 r^2} = \\
p_3^2\int~d^4r~\oint_C dz_2~dz_3  {\cal{M}}^{(n)} (z_2,z_3) 
\le\frac{(p_1+r)^2}{p_3^2}\ri^{z_2} \le\frac{(p_2-r)^2}{p_3^2}\ri^{z_3}
\frac{1}{(p_1 + r)^2 (p_2 - r)^2 r^2} = \\
\int~d^4r~\oint_C dz_2~dz_3  {\cal{M}}^{(n)} (z_2,z_3) 
\frac{1}{((p_1 + r)^2)^{1-z_2} ((p_2 - r)^2)^{1-z_3} r^2}\frac{1}{(p_3^2)^{z_2+z_3-1}} = \\
\pi^2\oint_C dz_2~dz_3  {\cal{M}}^{(n)} (z_2,z_3) J(1-z_3,1-z_2,1)\frac{1}{(p_3^2)^{z_2+z_3-1}} = \\
\pi^2\oint_C dz_2~dz_3~du~dv~ x^u y^v {\cal{M}}^{(n)} (z_2,z_3) D^{(u,v)}[1-z_3,1-z_2,1], 
\end{eqnarray*}
where $D^{(u,v)}[1-z_3,1-z_2,1]$ is defined above as the MB transform of the corresponding three-point 
integral, according to description done in the previous subsection \footnote{Factor $\pi^2$ appears due to usage of the original measure 
$d^4r$ in the momentum space.}.

Thus, one can derive the first recursive formula for the MB transforms, which relates the transforms of 
two neighbour UD functions \cite{Usyukina:1992jd,Usyukina:1993ch}
\begin{eqnarray*}
{\cal{M}}^{(n+1)} (u,v) = \pi^2 \oint_C dz_2~dz_3 {\cal{M}}^{(n)} (z_2,z_3)~ D^{(u,v)}[1-z_3,1-z_2,1].
\end{eqnarray*}
 
The function  $D^{(u,v)}[1-z_3,1-z_2,1]$ is a nontrivial combination of the Euler gamma 
functions in numerator and denominator,  
\begin{eqnarray*}
D^{(u,v)}[1-z_3,1-z_2,1] = \frac{\G \le z_2 - u \ri \G \le z_3 - v\ri \G \le 1 - z_2 - z_3 + u + v  \ri} 
{\G \le 1 +  z_2 + z_3 \ri \G \le 1 - z_2 \ri \G \le  1- z_3\ri} 
\G \le -u \ri \G \le -v \ri \G \le 1 + u + v \ri.
\end{eqnarray*}
This formula is written from definition (\ref{J-arb}).

\section{Reduction of the multi-fold MB transforms}

The formulas of the previous subsection are in some sense consequence of the ladder-like topology of the diagram.
We did not make any integration in the complex planes of MB parameters.   

Consider the diagram on the l.h.s. of fig. (\ref{figure-9}). We can repeat the trick of the previous section 
and obtain the following MB representation for it 

\begin{eqnarray}
\frac{1}{p_3^2}\int~Dr~\oint dz_2dz_3\le\frac{(p_1+r)^2}{p_3^2}\ri^{z_2} \le\frac{(p_2-r)^2}{p_3^2}\ri^{z_3}
\frac{D^{(z_2,z_3)}[1+\ve_2,1+\ve_1,1+\ve_3]}{[(p_1 + r)^2]^{1+\ve_2} [(p_2 - r)^2]^{1+\ve_1} [r^2]^{1+\ve_3}} = \no\\
\frac{1}{p_3^2}\oint_C~Dr~dz_2dz_3\frac{1}{[p_2^3]^{z_2+z_3}}
\frac{D^{(z_2,z_3)}[1+\ve_2,1+\ve_1,1+\ve_3]}{[(p_1 + r)^2]^{1+\ve_2-z_2} [(p_2 - r)^2]^{1+\ve_1-z_3} [r^2]^{1+\ve_3}} 
= \no\\
\frac{1}{p_3^2}\oint_C~dz_2dz_3\frac{1}{[p_3^2]^{z_2+z_3}}J(1+\ve_1-z_3,1+\ve_2-z_2,1+\ve_3)D^{(z_2,z_3)}[1+\ve_2,1+\ve_1,1+\ve_3] = \no\\
\frac{1}{(p_3^2)^2}\oint_C~dz_2dz_3~du~dv~x^u~y^v D^{(u,v)}[1+\ve_1-z_3,1+\ve_2-z_2,1+\ve_3] \times \no\\
D^{(z_2,z_3)}[1+\ve_2,1+\ve_1,1+\ve_3] \label{Tr-U-2}
\end{eqnarray}

This formula is derived in analogy with the MB transforms of the UD functions in the previous section. 
In detail, the procedure looks like follows. First, we calculate 
the MB transform of the leftmost triangle integral, this yields another triangle integral with indices depending on 
the complex variables of the previous MB transform. This procedure will be used in all the constructions below. 

The general strategy of the present investigation consists in the expanding of the r.h.s.  in terms 
of $\ve_i.$  The coefficients of this expansion will contain $\psi$ functions and its derivatives.  In the r.h.s. on fig. (\ref{figure-9}) 
we cannot put all the values of   $\ve_i$ immediately equal to zero without  expanding these r.h.s.  in terms of $\ve_i$ 
and observing that poles in  $\ve_i$ disappear. Thus, instead of a Laurent series we obtain a
Taylor series. Comparing the coefficients in front of the different products of powers of $\ve_i$  
we can derive an infinite number of new relations for the two-fold MB integrals over the complex variables $z_2$ and $z_3.$

\subsection{Description of the momentum integral}

Notation to be used
$$\int_n(\ve_1,\ve_2,\ve_3)$$   
corresponds to the momentum integral with incoming momenta $p_1,p_2,p_3$ as they are depicted on the l.h.s. of 
Fig.(\ref{figure-9}), with the integral measure defined in (\ref{k-measure}), but with  $n$ loops\footnote{with amputated external legs}.  
The $\ve$-terms  $\ve_1,\ve_2,\ve_3$ appear in the indices of lines for the first eight propagators on the left side of the diagram, namely, in the 
indices  $1+\ve_1,1+\ve_2,1+\ve_1$ on the upper side of the diagram, $1+\ve_2,1+\ve_1,1+\ve_2$ on the lower side of the diagram, and $1+\ve_3$ on the first 
two rungs. The rest of lines have indices equal to 1. These positions and values of the indices are indicated in the third 
line of Fig.(\ref{figure-1}) (in the position space).

\subsection{Two-fold MB transform for the two-rung ladder}

As a first step  to our formulas, we reproduce Eq.(25) of Ref.\cite{Usyukina:1992jd}. This is already done 
in Section 2, however we make it again just to introduce the notation that will be used for the higher 
rung diagrams.

We consider a diagrammatic relation that can be obtained by integral convolution of the leftmost external 
point of the diagrams in the first line of  fig. (\ref{figure-1}) with the line that has index $2-\ve_3.$ 
In such a way the Dirac $\delta$-function is produced which eliminates one of the integrations 
and the leftmost point on the l.h.s. is converted to the external vertex, that is, it is not a vertex of 
integration longer. The formula which can be used for this purpose is identity (\ref{remove}).

Then, we take the Fourier transform of each one of the diagrams in the diagrammatic relation 
obtained in that way. By keeping all the factors that appear after making the Fourier transform on 
both the parts of this new diagrammatic equation, we come to 
\begin{eqnarray*}
J^3 \frac{\G(1+\ve_3)}{\G(1-\ve_3)}\frac{1}{(p_1^2)^{1-\ve_1}(p_2^2)^{1-\ve_2}}\int_2(-\ve_1,-\ve_2,-\ve_3) =  \\
J\left[ \frac{J}{\ve_2\ve_3} \frac{\G(1+\ve_3)}{\G(1-\ve_3)} 
\frac{1}{(p_1^2)^{1-\ve_1} p_2^2 (p_3^2)^{1+\ve_3}}  J(1,1,1+\ve_1)  + \right. \\  
\left.  \frac{J}{\ve_1\ve_2} \frac{\G(1+\ve_3)}{\G(1-\ve_3)} 
\frac{1}{(p_1^2)^{1-\ve_1} (p_2^2)^{1-\ve_2} (p_3^2)^{1+\ve_3}}  J(1,1,1-\ve_3)  + \right. \\
\left. \frac{J}{\ve_1\ve_3} \frac{\G(1+\ve_3)}{\G(1-\ve_3)} 
\frac{1}{p_1^2 (p_2^2)^{1-\ve_2}  (p_3^2)^{1+\ve_3}}  J(1,1,1+\ve_2)  \right].
\end{eqnarray*}
After a simple algebra we derive an equation
\begin{eqnarray*}
\int_2(-\ve_1,-\ve_2,-\ve_3) = \frac{J^{-1}}{(p_3^2)^{1+\ve_3}}
\left[ \frac{1}{\ve_2\ve_3} \frac{1}{(p_2^2)^{\ve_2}}  J(1,1,1+\ve_1)  + \right. \\  
\left.  \frac{1}{\ve_1\ve_2}   J(1,1,1-\ve_3) + \frac{1}{\ve_1\ve_3} \frac{1}{(p_1^2)^{\ve_1}}  J(1,1,1+\ve_2)
\right],
\end{eqnarray*}
from which by changing the signs of all the values $\ve_i$ we obtain  
\begin{eqnarray}
\int_2(\ve_1,\ve_2,\ve_3) = \frac{J}{(p_3^2)^{1-\ve_3}}
\left[ \frac{1}{\ve_2\ve_3} \frac{1}{(p_2^2)^{-\ve_2}}  J(1,1,1-\ve_1)  + \right. \no\\  
\left.  \frac{1}{\ve_1\ve_2}   J(1,1,1+\ve_3) + \frac{1}{\ve_1\ve_3} \frac{1}{(p_1^2)^{-\ve_1}}  J(1,1,1-\ve_2)
\right].            \label{25U}
\end{eqnarray}
This is exactly Eq.(25) of Ref.\cite{Usyukina:1992jd}. This equation has been derived diagrammatically in the 
previous section. In that case the Fourier transform of the first line of fig. (\ref{figure-1}) has been done. 
This approach to reproduce Eq.(25) of Ref.\cite{Usyukina:1992jd} can be considered as a cross-check of 
the procedure used in the previous section to produce fig. (\ref{figure-9}).

At this moment we start to use the matter developed in the previous section. First of all, we rewrite 
the r.h.s. of Eq.(\ref{25U}) in the MB transformed representation 
\begin{eqnarray}
\int_2(\ve_1,\ve_2,\ve_3) = \frac{J}{(p_3^2)^{1-\ve_3}}
\left[ \frac{1}{\ve_2\ve_3} \frac{ (p_2^2)^{\ve_2} }{(p_3^2)^{1-\ve_1}} 
\oint_C du~dv~x^u~y^v D^{(u,v)}[1-\ve_1] + \right.\no\\ 
\left. \frac{1}{\ve_1\ve_2} \frac{ 1 }{(p_3^2)^{1+\ve_3}} \oint_C du~dv~x^u~y^v D^{(u,v)}[1+\ve_3] + 
\frac{1}{\ve_1\ve_3} \frac{ (p_1^2)^{\ve_1} }{(p_3^2)^{1-\ve_2}} 
\oint_C du~dv~x^u~y^v D^{(u,v)}[1-\ve_2] \right], \label{25U2}
\end{eqnarray}
where we have introduced a notation 
\begin{eqnarray*}
D^{(u,v)}[1+\nu] \equiv  D^{(u,v)}[1,1,1+\nu].
\end{eqnarray*}
Eq.(\ref{25U2}) can be written as 
\begin{eqnarray}
\int_2(\ve_1,\ve_2,\ve_3) = \frac{J}{(p_3^2)^{2}}
\left[ \frac{1}{\ve_2\ve_3} \le\frac{ p_2^2}{p_3^2}\ri^{\ve_2} 
\oint_C du~dv~x^u~y^v D^{(u,v)}[1-\ve_1] + \right.\no\\ 
\left. \frac{1}{\ve_1\ve_2} \oint_C du~dv~x^u~y^v D^{(u,v)}[1+\ve_3] + 
\frac{1}{\ve_1\ve_3}   \le\frac{ p_1^2}{p_3^2}\ri^{\ve_1} 
\oint_C du~dv~x^u~y^v D^{(u,v)}[1-\ve_2] \right] = \no\\
\frac{J}{(p_3^2)^{2}}
\left[ \frac{1}{\ve_2\ve_3} \oint_C du~dv~x^u~y^{v+\ve_2} D^{(u,v)}[1-\ve_1] + \right.\no\\ 
\left. \frac{1}{\ve_1\ve_2} \oint_C du~dv~x^u~y^v D^{(u,v)}[1+\ve_3] + 
\frac{1}{\ve_1\ve_3}  
\oint_C du~dv~x^{u+\ve_1}~y^v D^{(u,v)}[1-\ve_2] \right] = \no\\
\frac{J}{(p_3^2)^{2}}
\oint_C du~dv~x^u~y^{v} \left[ \frac{D^{(u,v-\ve_2)}[1-\ve_1]}{\ve_2\ve_3}
+  \frac{D^{(u,v)}[1+\ve_3]}{\ve_1\ve_2}  + \frac{ D^{(u-\ve_1,v)}[1-\ve_2]}{\ve_1\ve_3}  \right] \equiv \no\\
\frac{1}{(p_3^2 )^{2}}\oint_C du~dv~x^u~y^{v} M_2^{(u,v)}[\ve_1,\ve_2,\ve_3]. \label{ssylka}
\end{eqnarray}
Here we shift the variable of integration in the complex plane. That means, the contour of integration $C$ 
still passes between the left and the right poles. Positions of the poles in the plane of two complex variables 
will be changed with that trick but their nature (left or right) remains unchanged.

On the other hand, we have formula (\ref{Tr-U-2}), from which we obtain  
\begin{eqnarray}
\oint_C~dz_2dz_3~D^{(u,v)}[1+\ve_1-z_3,1+\ve_2-z_2,1+\ve_3] 
D^{(z_2,z_3)}[1+\ve_2,1+\ve_1,1+\ve_3] =  \no\\
J\left[ \frac{D^{(u,v-\ve_2)}[1-\ve_1]}{\ve_2\ve_3}
+  \frac{D^{(u,v)}[1+\ve_3]}{\ve_1\ve_2}  + \frac{ D^{(u-\ve_1,v)}[1-\ve_2]}{\ve_1\ve_3}  \right] = 
M_2^{(u,v)}[\ve_1,\ve_2,\ve_3].  \label{implicit-1}
\end{eqnarray}
This formula is valid for any $u,v$ and presents by itself a nontrivial result which can be used in practical
applications of the MB integration. This is the two-fold MB transform of two-loop integral, and it is valid not only 
in the limit of all $\ve_i \rightarrow 0$ but for any  $\ve_i.$ Expanding this formula in terms of 
$\ve_i$ we obtain infinite number of new relations. The explicit form of  $M_2^{(u,v)}[\ve_1,\ve_2,\ve_3]$ 
is calculated in the next section in the limit of vanishing $\ve_i,$ the result is Eq. (\ref{nah}), 
\begin{eqnarray*}
\oint_C~dz_2dz_3~D^{(u,v)}[1-z_3,1-z_2,1] D^{(z_2,z_3)}[1,1,1] =  \no\\
\G^2(1 + u + v)\G^2(-u)\G^2(-v) \times\no\\
\left[\frac{1}{2}\le\psi'(-v) + \psi'(-u)\ri - \frac{1}{2}\le\psi(-v) - \psi(-u)\ri^2 
- \frac{3}{2}\le\G(1-\ve)\G(1+ \ve) \ri^{(2)}_{\ve = 0} \right].
\end{eqnarray*}

\subsection{Two-fold MB transform for the three-rung ladder}

We can derive a new formula for the two-fold MB transformation going to higher loops in the ladder diagrams. 
We start to work with integral $\int_3(\ve_1,\ve_2,\ve_3)$ which appears on the l.h.s. in the third line of
Fig.(\ref{figure-1}) and take the Fourier transform of each one of the diagrams of that line. The Fourier 
transform of the l.h.s. can be written as 
\begin{eqnarray*}
\le \frac{\G(1-\ve_1)}{\G(1+\ve_1)} \ri^3 \le \frac{\G(1-\ve_2)}{\G(1+\ve_2)} \ri^3 
\le \frac{\G(1-\ve_3)}{\G(1+\ve_3)} \ri^2
\frac{1}{p_1^2}\frac{1}{p_2^2} \int_3(-\ve_1,-\ve_2,-\ve_3)
\end{eqnarray*}
and the whole diagrammatic relation becomes as 
\begin{eqnarray*}
J^3 \frac{\G(1+\ve_3)}{\G(1-\ve_3)}\frac{1}{p_1^2p_2^2}\int_3(-\ve_1,-\ve_2,-\ve_3) = 
J^2 \frac{1}{\ve_2\ve_3} \frac{\G(1+\ve_3)}{\G(1-\ve_3)} 
\frac{1}{(p_1^2)^{1-\ve_1} p_2^2 (p_3^2)^{1-\ve_2}} \int_2(-\ve_1)  +   \\  
\frac{J^3}{\ve_1\ve_2} \frac{\G(1+\ve_3)}{\G(1-\ve_3)} 
\frac{1}{(p_1^2)^{1-\ve_1} (p_2^2)^{1-\ve_2} p_3^2} \int_2(-\ve_1,-\ve_2,-\ve_3)   +  
\frac{J^2}{\ve_1\ve_3} \frac{\G(1+\ve_3)}{\G(1-\ve_3)} 
\frac{1}{p_1^2 (p_2^2)^{1-\ve_2}  (p_3^2)^{1-\ve_1}} \int_2(-\ve_2). 
\end{eqnarray*}
We kept all the factors that appear after making the Fourier transform on both the parts of the
diagrammatic equation. Here we use a brief notation 
\begin{eqnarray*}
\int_n(-\ve_1) \equiv \int_n(-\ve_1,0, \ve_1), ~~~~ \int_n(-\ve_2) \equiv \int_n(0,-\ve_2, \ve_2).
\end{eqnarray*}
After a little algebra we obtain an expansion  of the three-loop integral in terms of the two-loop integrals, 
\begin{eqnarray*}
\int_3(-\ve_1,-\ve_2,-\ve_3) = J^{-1} \left[\frac{1}{(p_1^2)^{-\ve_1} (p_3^2)^{1-\ve_2}}
\frac{1}{\ve_2\ve_3} \int_2(-\ve_1) + \right.\\
\left. \frac{J}{(p_1^2)^{-\ve_1} (p_2^2)^{-\ve_2} p_3^2}
\frac{1}{\ve_1\ve_2}\int_2(-\ve_1,-\ve_2,-\ve_3)  + 
\frac{1}{(p_2^2)^{-\ve_2} (p_3^2)^{1-\ve_1} }  \frac{1}{\ve_1\ve_3} \int_2(-\ve_2) \right]. 
\end{eqnarray*}
By changing the signs of all values $\ve_i$ we obtain an equation 
\begin{eqnarray*}
\int_3(\ve_1,\ve_2,\ve_3) = J \left[\frac{1}{(p_1^2)^{\ve_1} (p_3^2)^{1+\ve_2}}
\frac{1}{\ve_2\ve_3} \int_2(\ve_1) + \right.\\
\left. \frac{J^{-1}}{(p_1^2)^{\ve_1} (p_2^2)^{\ve_2} p_3^2}
\frac{1}{\ve_1\ve_2}\int_2(\ve_1,\ve_2,\ve_3)  + 
\frac{1}{(p_2^2)^{\ve_2} (p_3^2)^{1+\ve_1} }  \frac{1}{\ve_1\ve_3} \int_2(\ve_2) \right].
\end{eqnarray*}
The previous equation can be re-written for the MB transforms, and we obtain the result 
\begin{eqnarray*}
\int_3(\ve_1,\ve_2,\ve_3) = J
\left[ \frac{1}{\ve_2\ve_3} \frac{1}{(p_1^2)^{\ve_1} (p_3^2)^{3 +\ve_2}} \oint_C du~dv~x^u~y^v M_2^{(u,v)}(\ve_1) 
+ \right. \\
\left. J^{-1}\frac{1}{\ve_1\ve_2} \frac{1}{(p_1^2)^{\ve_1} (p_2^2)^{\ve_2} (p_3^2)^{3}} 
\oint_C du~dv~x^u~y^v M_2^{(u,v)}(\ve_1,\ve_2,\ve_3) + \right.\\
\left. \frac{1}{\ve_1\ve_3} \frac{1}{(p_2^2)^{\ve_2} (p_3^2)^{3+\ve_1}} \oint_C du~dv~x^u~y^v M_2^{(u,v)}(\ve_2)
\right] =  \\
J
\left[ \frac{1}{\ve_2\ve_3} \le\frac{p_3^2}{p_1^2}\ri^{\ve_1} \frac{1}{(p_3^2)^{3-\ve_3}} 
\oint_C du~dv~x^u~y^v M_2^{(u,v)}(\ve_1) + \right. \\
\left. 
J^{-1}\frac{1}{\ve_1\ve_2} \le\frac{p_3^2}{p_1^2}\ri^{\ve_1} \le\frac{p_3^2}{p_2^2}\ri^{\ve_2}
\frac{1}{(p_3^2)^{3-\ve_3}} 
\oint_C du~dv~x^u~y^v M_2^{(u,v)}(\ve_1,\ve_2,\ve_3) \right.\\
\left. + \frac{1}{\ve_1\ve_3} \le\frac{p_3^2}{p_2^2}\ri^{\ve_2} \frac{1}{(p_3^2)^{3-\ve_3}} 
\oint_C du~dv~x^u~y^v M_2^{(u,v)}(\ve_2)\right] = \\
\frac{J}{(p_3^2)^{3-\ve_3}}
\left[ \frac{1}{\ve_2\ve_3} \oint_C du~dv~x^{u-\ve_1}~y^v M_2^{(u,v)}(\ve_1) +  
J^{-1}\frac{1}{\ve_1\ve_2} \oint_C du~dv~x^{u-\ve_1}~y^{v-\ve_2} M_2^{(u,v)}(\ve_1,\ve_2,\ve_3) \right.\\
\left. + \frac{1}{\ve_1\ve_3}  \oint_C du~dv~x^u~y^{v-\ve_2} M_2^{(u,v)}(\ve_2)\right] = \\
\frac{J}{(p_3^2)^{3-\ve_3}}
\oint_C du~dv~x^u~y^v \left[\frac{1}{\ve_2\ve_3} M_2^{(u+\ve_1,v)}(\ve_1) +  
\frac{J^{-1}}{\ve_1\ve_2} M_2^{(u+\ve_1,v+\ve_2)}(\ve_1,\ve_2,\ve_3) + \right.\\
\left. + \frac{1}{\ve_1\ve_3} M_2^{(u,v+\ve_2)}(\ve_2)\right] 
\equiv 
\frac{1}{(p_3^2)^{3-\ve_3}}\oint_C du~dv~x^u~y^v  M_3^{(u,v)}(\ve_1,\ve_2,\ve_3).
\end{eqnarray*}
Here we have shifted the variable of integration in the complex plane. That means, the contour of integration $C$ 
still passes between the left and the right poles. Positions of the poles in the planes of two complex variables 
will be changed but their nature (to belong to the set of left poles or to the set of the right poles) 
cannot be changed with such a trick. 

On the other side, in complete analogy with Eq.(\ref{Tr-U-2}) we obtain 
\begin{eqnarray}
\int_3 (\ve_1,\ve_2,\ve_3) = \no\\
\frac{1}{(p_3^2)^2}
\int~Dr~\oint dz_2dz_3\le\frac{(p_1+r)^2}{p_3^2}\ri^{z_2} \le\frac{(p_2-r)^2}{p_3^2}\ri^{z_3}
\frac{M_2^{(z_2,z_3)}(\ve_1,\ve_2,\ve_3)}{[(p_1 + r)^2]^{1+\ve_1} [(p_2 - r)^2]^{1+\ve_2} r^2} = \no\\
\frac{1}{(p_3^2)^2} \int~Dr~\oint_C~dz_2dz_3\frac{1}{[p_3^2]^{z_2+z_3}}
\frac{M_2^{(z_2,z_3)}(\ve_1,\ve_2,\ve_3)}{[(p_1 + r)^2]^{1+\ve_1-z_2} [(p_2 - r)^2]^{1+\ve_2-z_3} r^2} = \no\\
\oint_C~dz_2dz_3\frac{1}{[p_3^2]^{2+z_2+z_3}}J(1+\ve_2-z_3,1+\ve_1-z_2, 1) M_2^{(z_2,z_3)}(\ve_1,\ve_2,\ve_3) = \no\\
\frac{1}{(p_3^2)^{3-\ve_3}}\oint_C~dz_2dz_3~dudv~x^uy^vD^{(u,v)}[1+\ve_2-z_3,  1+\ve_1-z_2, 1]
M_2^{(z_2,z_3)}(\ve_1,\ve_2,\ve_3). \label{Tr-U-3}
\end{eqnarray}
Thus, we derive that
\begin{eqnarray}
 M_3^{(u,v)}[\ve_1,\ve_2,\ve_3] = \oint_C~dz_2dz_3 D^{(u,v)}[1+\ve_2-z_3,1+\ve_1-z_2,1]M_2^{(z_2,z_3)}(\ve_1,\ve_2,\ve_3) = \no\\
J \left[\frac{1}{\ve_2\ve_3} M_2^{(u+\ve_1,v)}(\ve_1) +  
\frac{J^{-1}}{\ve_1\ve_2} M_2^{(u+\ve_1,v+\ve_2)}(\ve_1,\ve_2,\ve_3) + 
\frac{1}{\ve_1\ve_3} M_2^{(u,v+\ve_2)}(\ve_2) \right].  \label{implicit-2}
\end{eqnarray}
This formula is valid for any $u,v$ and for any  $\ve_1,\ve_2.$  Expanding this formula in terms of 
$\ve_i$ we can obtain infinite number of new integral relations. In the next section we calculate the limit $\ve_i \rightarrow 0$ of 
$M_3^{(u,v)}[\ve_1,\ve_2,\ve_3].$

\subsection{Two-fold MB transform for the four-rung ladder}

In the next paragraphs of this section we consider the case of the four-loop momentum triangle ladder diagram.
It looks like the diagram on the l.h.s. of the third line of Fig.(\ref{figure-1}) but with one more rung. 
To get that graphical representation, 
the diagrams in the third line of  fig. (\ref{figure-1}) are integrated with three more propagators, 
the index of each propagator  is equal to 1.  The Fourier transform of the l.h.s. of the diagrammatic relation
obtained in such a way contains  the integral $\int_4(\ve_1,\ve_2,\ve_3),$ 
\begin{eqnarray*}
\le \frac{\G(1-\ve_1)}{\G(1+\ve_1)} \ri^3 \le \frac{\G(1-\ve_2)}{\G(1+\ve_2)} \ri^3 
\le \frac{\G(1-\ve_3)}{\G(1+\ve_3)} \ri^2
\frac{1}{p_1^2}\frac{1}{p_2^2} \int_4(-\ve_1,-\ve_2,-\ve_3), 
\end{eqnarray*}
and we derive the identity 
\begin{eqnarray*}
J^3 \frac{\G(1+\ve_3)}{\G(1-\ve_3)}\frac{1}{p_1^2p_2^2}\int_4(-\ve_1,-\ve_2,-\ve_3) = 
J^2 \frac{1}{\ve_2\ve_3} \frac{\G(1+\ve_3)}{\G(1-\ve_3)} 
\frac{1}{p_1^2 p_2^2 (p_3^2)^{1-\ve_2}} \int_3(-\ve_1)  +   \\  
\frac{J^3}{\ve_1\ve_2} \frac{\G(1+\ve_3)}{\G(1-\ve_3)} 
\frac{1}{p_1^2 p_2^2 p_3^2} \int_3(-\ve_1,-\ve_2,-\ve_3)   +  
\frac{J^2}{\ve_1\ve_3} \frac{\G(1+\ve_3)}{\G(1-\ve_3)} 
\frac{1}{p_1^2 p_2^2  (p_3^2)^{1-\ve_1}} \int_3(-\ve_2). 
\end{eqnarray*}
After a little algebra we obtain an expansion  of the four-loop integral in terms of the three-loop integrals, 
\begin{eqnarray*}
\int_4(-\ve_1,-\ve_2,-\ve_3) = J^{-1} \left[\frac{1}{(p_3^2)^{1-\ve_2}}
\frac{1}{\ve_2\ve_3} \int_3(-\ve_1) + 
\frac{J}{p_3^2} \frac{1}{\ve_1\ve_3}\int_3(-\ve_1,-\ve_2,-\ve_3)  + \right. \\
\left. \frac{1}{(p_3^2)^{1-\ve_1} }  \frac{1}{\ve_1\ve_3} \int_3(-\ve_2) \right],
\end{eqnarray*}
from which by changing the signs of all the values $\ve_i$ we obtain an equation 
\begin{eqnarray*}
\int_4(\ve_1,\ve_2,\ve_3) = J \left[\frac{1}{(p_3^2)^{1+\ve_2}}\frac{1}{\ve_2\ve_3} \int_3(\ve_1) + 
\frac{J^{-1}}{p_3^2} \frac{1}{\ve_1\ve_3}\int_3(\ve_1,\ve_2,\ve_3)  + \right. \\
\left. \frac{1}{(p_3^2)^{1+\ve_1} }  \frac{1}{\ve_1\ve_3} \int_3(\ve_2) \right].
\end{eqnarray*}
This equation can be re-written for the MB transforms 
\begin{eqnarray*}
\int_4(\ve_1,\ve_2,\ve_3) = J
\left[ \frac{1}{\ve_2\ve_3} \frac{1}{ (p_3^2)^{1+\ve_2} (p_3^2)^{3+\ve_1} } \oint_C du~dv~x^u~y^v M_3^{(u,v)}(\ve_1) 
+ \right. \\
\left. J^{-1}\frac{1}{\ve_1\ve_2} \frac{1}{(p_3^2)^{4-\ve_3}} 
\oint_C du~dv~x^u~y^v M_3^{(u,v)}(\ve_1,\ve_2,\ve_3) + \right.\\
\left. \frac{1}{\ve_1\ve_3} \frac{1}{(p_3^2)^{1+\ve_1} (p_3^2)^{3+\ve_2}} \oint_C du~dv~x^u~y^v 
M_3^{(u,v)}(\ve_2)\right] =  \\
\frac{J}{(p_3^2)^{4-\ve_3}}
\oint_C du~dv~x^u~y^v \left[\frac{1}{\ve_2\ve_3} M_3^{(u,v)}(\ve_1) +  
\frac{J^{-1}}{\ve_1\ve_2} M_3^{(u,v)}(\ve_1,\ve_2,\ve_3) + \frac{1}{\ve_1\ve_3} M_3^{(u,v)}(\ve_2)\right] \equiv \\
\frac{1}{(p_3^2)^{4-\ve_3}}\oint_C du~dv~x^u~y^{v} M_4^{(u,v)}[\ve_1,\ve_2,\ve_3].
\end{eqnarray*}
On the other hand, in analogy to Eq.(\ref{Tr-U-2}),   
\begin{eqnarray}
\int_4 (\ve_1,\ve_2,\ve_3) = \no\\
\frac{1}{(p_3^2)^{3-\ve_3}}
\int~Dr~\oint dz_2dz_3\le\frac{(p_1+r)^2}{p_3^2}\ri^{z_2} \le\frac{(p_2-r)^2}{p_3^2}\ri^{z_3}
\frac{M_3^{(z_2,z_3)}(\ve_1,\ve_2,\ve_3)}{(p_1 + r)^2 (p_2 - r)^2 r^2} = \no\\
\frac{1}{(p_3^2)^{3-\ve_3}} \int~Dr~\oint_C~dz_2dz_3\frac{1}{[p_3^2]^{z_2+z_3}}
\frac{M_3^{(z_2,z_3)}(\ve_1,\ve_2,\ve_3)}{[(p_1 + r)^2]^{1-z_2} [(p_2 - r)^2]^{1-z_3} r^2} = \no\\
\frac{1}{(p_3^2)^{3-\ve_3}}\oint_C~dz_2dz_3\frac{1}{[p_3^2]^{z_2+z_3}}J(1-z_3,1-z_2,1) M_3^{(z_2,z_3)}
(\ve_1,\ve_2,\ve_3) = \no\\
\frac{1}{(p_3^2)^{4-\ve_3}}\oint_C~dz_2dz_3~dudv~x^uy^vD^{(u,v)}[1 -z_3,1-z_2,1]
M_3^{(z_2,z_3)}(\ve_1,\ve_2,\ve_3). \label{Tr-U-4}
\end{eqnarray}
We obtain 
\begin{eqnarray}
\oint_C~dz_2dz_3 D^{(u,v)}[1-z_3,1-z_2,1]M_3^{(z_2,z_3)}(\ve_1,\ve_2,\ve_3) = \no\\
J \left[\frac{1}{\ve_2\ve_3} M_3^{(u,v)}(\ve_1) +  
\frac{J^{-1}}{\ve_1\ve_2} M_3^{(u,v)}(\ve_1,\ve_2,\ve_3) + 
\frac{1}{\ve_1\ve_3} M_3^{(u,v)}(\ve_2) \right] = M_4^{(u,v)}[\ve_1,\ve_2,\ve_3]. \label{implicit-3}
\end{eqnarray}
New integral relations can be derived by expanding this formula in terms of $\ve_i,$ and this formula is valid for any $u,v.$

\section{One-fold MB transforms}

In this section we collect the useful formulas for the one-fold MB transforms,

\begin{eqnarray*}
\oint_C~dz~x^{z}\G^2(-z)\G^2(1+z) = - \frac{\ln x}{1-x}, \\
\oint_{C} dz~ x^{z} \left\{\G^2 \le -z \ri \G \le z + 1 \ri \G^* \le z \ri \right\} = -\le\frac{1}{2}\ln^2x + 2\zeta(2)\ri + 
\ln(1-x)\ln x + {\rm Li}_2(x), \\
 \oint_C~dz~x^{z}\G^2(-z)\G^2(1+z) \psi(1+z) = \\
\frac{1}{(1-x)}\left[-\psi(1)\ln x + \ln(1-x) \ln x -\zeta(2) + {\rm Li}_2(x) \right].
\end{eqnarray*}

It is possible to work with the MB transforms by making a use of such tricks like derivation with respect to 
parameter and  integration by parts in the complex plane.  
Here we demonstrate several examples. First of all, after the derivation with respect to $x$ of 
\begin{eqnarray*}
\frac{1}{1+x} = \oint_C~dz~x^{z}\G(-z)\G(1+z)  
\end{eqnarray*}
we obtain 
\begin{eqnarray*}
\frac{x}{(1+x)^2} = \oint_C~dz~x^{z}\G(1-z)\G(1+z).  
\end{eqnarray*}
The result can be checked directly by the counting of the residues. Now, let us check that the integral of the total derivative
is equal to zero. Indeed, we have 
\begin{eqnarray}
0 = \oint_C~dz~\frac{d}{dz}\{x^{z}\G(-z)\G(1+z)\} = \ln x \oint_C~dz~x^{z}\G(-z)\G(1+z) + \no\\
\oint_C~dz~x^z\G(-z)\G(1+z)(-\psi(-z) + \psi(1+z)) = \no\\
 \frac{\ln x}{1+x} - \oint_C~dz~x^{z}\frac{\G(-z)\G(1+z)}{z} + \oint_C~dz~x^z\G(-z)\G(1+z)(\psi(1+z) - \psi(1-z)). \label{res}
\end{eqnarray}
We calculate all the integrals on the r.h.s. explicitly and show that their sum is zero. The contribution  of residues leads to
\begin{eqnarray*}
\oint_C~dz~x^z\G(-z)\G(1+z)\psi(1+z) = \psi(1) - x\psi(2) + x^2\psi(3) + \dots, \\
\oint_C~dz~x^z\G(-z)\G(1+z)\psi(1-z) = \psi(1) - x\psi(1) + x^2\psi(2) - x^3\psi(3) + \dots - \ln x + \frac{\ln x}{1+x},\\
 \oint_C~dz~x^{z}\frac{\G(-z)\G(1+z)}{z} = \ln x -\ln(1+x).
\end{eqnarray*}
Substituting these results in (\ref{res}) we reproduce zero in the l.h.s. of it.

\section{Explicit results of the two-fold MB transform}

In order to use formulas obtained for the two-fold integration explicitly, we need to calculate the r.h.s. 
of the equations in the limit of vanishing $\ve_i.$  First, we consider the result for $\int_2(\ve_1,\ve_2,\ve_3).$
There are two ways to derive the result in this limit for the r.h.s.

\subsection{Belokurov-Usyukina expansion for the higher rung ladder}

According to Eqs. (\ref{ssylka}) and (\ref{J-arb-2}), we have to find the limit  
\begin{eqnarray*}
\lim_{\ve_2 \rar 0, \ve_1 \rar 0}  M_2^{(u,v)}(\ve_1,\ve_2,\ve_3) = \no\\
\lim_{\ve_2 \rar 0, \ve_1 \rar 0} J\left[\frac{D^{(u,v-\ve_2)}[1-\ve_1]}{\ve_2\ve_3}
+  \frac{D^{(u,v)}[1+\ve_3]}{\ve_1\ve_2}  + \frac{ D^{(u-\ve_1,v)}[1-\ve_2]}{\ve_1\ve_3}\right] = \no\\
\lim_{\ve_2 \rar 0, \ve_1 \rar 0} \left[
\frac{1}{\ve_1\ve_2}\frac{\G(-u)\G(-v)\G(\ve_1 + \ve_2 -u)\G(\ve_1 + \ve_2 -v)\G^2(1-\ve_1-\ve_2 + u + v)}
{\G(1-\ve_3)\G(1+\ve_3)} \right. \no\\
\left.
- \frac{1}{\ve_2(\ve_1+\ve_2)}\frac{\G(-u)\G(\ve_2-v)\G(\ve_1-u)\G(\ve_1 + \ve_2 -v)\G^2(1-\ve_1-\ve_2 + u + v)}
{\G(1-\ve_1)\G(1+\ve_1)} \right.\no\\
\left.
- \frac{1}{\ve_1(\ve_1+\ve_2)}\frac{\G(\ve_1-u)\G(-v)\G(\ve_1+\ve_2-u)\G(\ve_2 -v)\G^2(1-\ve_1-\ve_2 + u + v)}
 {\G(1-\ve_2)\G(1+\ve_2)} \right] = \no\\
\G^2(1 + u + v) \lim_{\ve_2 \rar 0, \ve_1 \rar 0} \left[
\frac{1}{\ve_1\ve_2}\frac{\G(-u)\G(-v)\G(\ve_1 + \ve_2 -u)\G(\ve_1 + \ve_2 -v)}{\G(1-\ve_3)\G(1+\ve_3)} \right. \no\\
\left.
- \frac{1}{\ve_2(\ve_1+\ve_2)}\frac{\G(-u)\G(\ve_2-v)\G(\ve_1-u)\G(\ve_1 + \ve_2 -v)}{\G(1-\ve_1)\G(1+\ve_1)} \right.\no\\
\left.
- \frac{1}{\ve_1(\ve_1+\ve_2)}\frac{\G(\ve_1-u)\G(-v)\G(\ve_1+\ve_2-u)\G(\ve_2 -v)}{\G(1-\ve_2)\G(1+\ve_2)} \right]. 
\end{eqnarray*}
After a simple calculus, the limit is 
\begin{eqnarray}
\lim_{\ve_1\rightarrow 0,\ve_2\rightarrow 0} M_2^{(u,v)}(\ve_1,\ve_2,\ve_3) = 
\G^2(1 + u + v)\G^2(-u)\G^2(-v) \times\no\\
\left[\frac{1}{2}\le\psi'(-v) + \psi'(-u)\ri - \frac{1}{2}\le\psi(-v) - \psi(-u)\ri^2 
- \frac{3}{2}\le\G(1-\ve)\G(1+ \ve) \ri^{(2)}_{\ve=0} \right]. \label{nah}
\end{eqnarray}
The last term is a value of Riemann zeta function multiplied by the first UD function. 
This result can be represented in other forms. For example, by  integrating by parts in the complex $u$ and $v$ 
planes we reproduce result (\ref{ssylka-3})  of the next subsection. However, this form is important since 
it stands in the right hand side of Eq. (\ref{implicit-2}) in the limit of vanishing $\ve_i.$

\subsection{Integration by parts in the complex plane}

Also, it can be helpful to work with the two-fold MB transforms by using integration by parts technique  in the complex plane which 
is described in the previous section. Indeed, as we have derived in Eq. (\ref{ssylka})

\begin{eqnarray}
\lim_{\ve_2 \rar 0, \ve_1 \rar 0} \oint_C du~dv~x^u~y^v M_2^{(u,v)}(\ve_1,\ve_2,\ve_3) = \no\\
\lim_{\ve_2 \rar 0, \ve_1 \rar 0} \left[\frac{1}{\ve_1\ve_2} \oint_C du~dv~x^u~y^v D^{(u,v)}[1+\ve_3] + \frac{1}{\ve_2\ve_3} y^{\ve_2} 
\oint_C du~dv~x^u~y^v D^{(u,v)}[1-\ve_1] + \right.\no\\
\left. 
\frac{1}{\ve_1\ve_3}   x^{\ve_1} \oint_C du~dv~x^u~y^v D^{(u,v)}[1-\ve_2] \right] = \no\\
\lim_{\ve_2 \rar 0, \ve_1 \rar 0} \oint_C du~dv~x^u~y^v J\Bigl[\frac{1}{\ve_2\ve_3} y^{\ve_2} D^{(u,v)}[1-\ve_1] + \frac{1}{\ve_1\ve_2} D^{(u,v)}[1+\ve_3] \Bigr.\no\\
\Bigl.
+ \frac{1}{\ve_1\ve_3} x^{\ve_1} D^{(u,v)}[1-\ve_2] \Bigr]. \label{not implicit-1}
\end{eqnarray}

From definition  (\ref{J-arb-2}), we write the integrand  explicitly
\begin{eqnarray*}
\frac{1}{\ve_1\ve_2}  D^{(u,v)}[1+\ve_3]  + \frac{1}{\ve_2\ve_3} y^{\ve_2} D^{(u,v)}[1-
\ve_1]  +  \frac{1}{\ve_1\ve_3} x^{\ve_1}D^{(u,v)}[1-\ve_1] = \\
\G(-u)\G(-v)\left[\frac{1}{\ve_1\ve_2}\frac{\G(\ve_1+\ve_2 -u)\G(\ve_1 + \ve_2 -v)\G^2(1-\ve_1 -\ve_2 +u+v)}{\G(1+ 
\ve_1 + \ve_2)\G(1- \ve_1 - \ve_2)} + \right.\\
\left. - \frac{1}{\ve_2(\ve_1+\ve_2)} y^{\ve_2}\frac{\G(\ve_1 - u)\G(\ve_1 - v)\G^2(1-\ve_1 + u + v)}
{\G(1+ \ve_1)\G(1- \ve_1)} \right.\\
\left. - \frac{1}{\ve_1(\ve_1+\ve_2)} x^{\ve_1}\frac{\G(\ve_2 -u)\G(\ve_2 -v)\G^2(1 - \ve_2 +u+v)}
{\G(1+ \ve_2)\G(1- \ve_2 )} \right].
\end{eqnarray*}
The brackets in this formula can be re-written in the following form 
\begin{eqnarray*}
\frac{1}{\ve_1\ve_2}\left[\frac{\G(\ve_1+\ve_2 -u)\G(\ve_1 + \ve_2 -v)\G^2(1-\ve_1 -\ve_2 +u+v)}{\G(1+ 
\ve_1 + \ve_2)\G(1- \ve_1 - \ve_2)} \right.\\
\left. - y^{\ve_2}\frac{\G(\ve_1 - u)\G(\ve_1 - v)\G^2(1-\ve_1 + u + v)}{\G(1+ \ve_1)\G(1- \ve_1)} \right] \\
 + \frac{1}{\ve_1^2}\left[\frac{\G(\ve_1 - u)\G(\ve_1 - v)\G^2(1-\ve_1 + u + v)}
{\G(1+ \ve_1)\G(1- \ve_1)} -  x^{\ve_1}\G(-u)\G(-v)\G^2(1 +u+v) \right] + o(\ve_2) = \\
\frac{1}{\ve_1} \G(-u)\G(-v)\G^2(1+u+v) \left[\psi(-u) + \psi(-v) -2\psi(1+u+v) - \ln y\right] \\
+ \le \frac{ \G(\ve_1 - u)\G(\ve_1 - v)\G^2(1-\ve_1 + u + v)}{\G(1+ \ve_1)\G(1- \ve_1)}\ri^{(2)}_{\ve_1=0} \\
- \ln y \le \G(\ve_1 - u)\G(\ve_1 - v)\G^2(1-\ve_1 + u + v)\ri'_{\ve_1=0}  \\
+ \frac{1}{\ve_1} \G(-u)\G(-v)\G^2(1+u+v) \left[\psi(-u) + \psi(-v) -2\psi(1+u+v) - \ln x \right]  \\
+ \frac{1}{2} \le \frac{\G(\ve_1 - u)\G(\ve_1 - v)\G^2(1-\ve_1 + u + v)}{\G(1+ \ve_1)\G(1- \ve_1)}\ri^{(2)}_{\ve_1=0} \\
- \frac{1}{2} \ln^2 x \G(-u)\G(-v)\G^2(1+u+v) + o(\ve_2) + o(\ve_1) . 
\end{eqnarray*}
First of all, we demonstrate that all the poles in $\ve_i$ disappear. 
The corresponding contribution taking into account the factor 
$\G(-u)\G(-v)$ has the form 
\begin{eqnarray*}
\frac{1}{\ve_1}\oint_C du~dv x^u y^v\G^2(-u)\G^2(-v)\G^2(1+u+v) \left[\le \psi(-u) + \psi(-v) -2\psi(1+u+v) - \ln y\ri \right.\\
+ \left. \le \psi(-u) + \psi(-v) -2\psi(1+u+v) - \ln x \ri \right] = \\
\frac{1}{\ve_1}\oint_C du~dv x^u y^v\G^2(-u)\G^2(-v)\G^2(1+u+v) \left[\le 2\psi(-u) - 2\psi(1+u+v) - \ln x\ri \right.\\
+ \left. \le  2\psi(-v) - 2\psi(1+u+v) - \ln y \ri \right] = \\
-\frac{1}{\ve_1}\oint_C du~dv (\pd_u + \pd_v)\{x^u y^v\G^2(-u)\G^2(-v)\G^2(1+u+v)\} = 0. 
\end{eqnarray*}
Thus, we have derived  
\begin{eqnarray*}
\lim_{\ve_2 \rar 0, \ve_1 \rar 0} \oint_C du~dv~x^u~y^v M_2^{(u,v)}(\ve_1,\ve_2,\ve_3) = \no\\
\oint_C du~dv~x^u~y^v \G(-u)\G(-v)\left[  \frac{3}{2}\le 
\frac{ \G(\ve_1 - u)\G(\ve_1 - v)\G^2(1-\ve_1 + u + v)}{\G(1+ \ve_1)\G(1- \ve_1)}\ri^{(2)}_{\ve_1=0} \right. \\
- \left. \ln y \le \G(\ve_1 - u)\G(\ve_1 - v)\G^2(1-\ve_1 + u + v)\ri'_{\ve_1=0} \right. \\
\left.
- \frac{1}{2} \ln^2 x \G(-u)\G(-v)\G^2(1+u+v)\right]. 
\end{eqnarray*}
As to the last two terms,  we can consider the integral transformations
\begin{eqnarray*}
\oint_C du~dv~x^u~y^v \G(-u)\G(-v)\left[- \ln y \le \G(\ve_1 - u)\G(\ve_1 - v)\G^2(1-\ve_1 + u + v)\ri'_{\ve_1=0} \right.\\
\left. - \frac{1}{2} \ln^2 x \G(-u)\G(-v)\G^2(1+u+v)\right]  = \\
- \ln y \oint_C du~dv~x^u~y^v \le \psi(-u) + \psi(-v) -2 \psi(1+u+v)\ri \G^2(-u)\G^2(-v)\G^2(1+u+v)\\
- \frac{1}{2} \ln^2 x \oint_C du~dv~x^u~y^v  \G^2(-u)\G^2(-v)\G^2(1+u+v)  = \\
\frac{1}{2} \ln y \oint_C du~dv~x^u~y^v \le \pd_u +\pd_v \ri \G^2(-u)\G^2(-v)\G^2(1+u+v)\\
- \frac{1}{2} \ln^2 x \oint_C du~dv~x^u~y^v  \G^2(-u)\G^2(-v)\G^2(1+u+v) = \\
-\frac{1}{2} \ln y (\ln x + \ln y)\oint_C du~dv~x^u~y^v \G^2(-u)\G^2(-v)\G^2(1+u+v)\\
- \frac{1}{2} \ln^2 x \oint_C du~dv~x^u~y^v  \G^2(-u)\G^2(-v)\G^2(1+u+v).
\end{eqnarray*}
Thus, the result for integral (\ref{not implicit-1}) is
\begin{eqnarray*}
\lim_{\ve_2 \rar 0, \ve_1 \rar 0} \oint_C du~dv~x^u~y^v M_2^{(u,v)}(\ve_1,\ve_2,\ve_3) = \no\\
\frac{3}{2}\oint_C du~dv~x^u~y^v \G(-u)\G(-v) \le 
\frac{ \G(\ve_1 - u)\G(\ve_1 - v)\G^2(1-\ve_1 + u + v)}{\G(1+ \ve_1)\G(1- \ve_1)}\ri^{(2)}_{\ve_1=0}  \no\\
 - \frac{1}{2}\le  \ln^2 x + \ln x \ln y  + \ln^2 y \ri \oint_C du~dv~x^u~y^v  \G^2(-u)\G^2(-v)\G^2(1+u+v) = \no\\
\frac{3}{2}\oint_C du~dv~x^u~y^v \G(-u)\G(-v) \le \G(\ve_1 - u)\G(\ve_1 - v)\G^2(1-\ve_1 + u + v)\ri_{\ve_1 =0}  \no\\
 - \frac{1}{2}\le  \ln^2 x + \ln x \ln y  + \ln^2 y \ri \oint_C du~dv~x^u~y^v  \G^2(-u)\G^2(-v)\G^2(1+u+v) \no\\
- \frac{3}{2}\le \G(1+ \ve_1)\G(1- \ve_1)\ri^{(2)}_{\ve_1=0} \oint_C du~dv~x^u~y^v \G^2(-u)\G^2(-v) \G^2(1+u+v). 
\end{eqnarray*}
This formula can be developed further and be transformed to the form without derivatives  
of the Euler $\psi$ function in the integrand,   
\begin{eqnarray*}
\oint_C du~dv~x^u~y^v \G(-u)\G(-v) \le \G(\ve_1 - u)\G(\ve_1 - v)\G^2(1-\ve_1 + u + v)
\ri^{(2)}_{\ve_1}  = \\ 
\oint_C du~dv~x^u~y^v \left[ -\pd_u \psi(-u) -\pd_v \psi(-v)  + \pd_u \psi(1+u+v) \right.\\
\left. + \pd_v \psi(1+u+v) + (\psi(-u) + \psi(-v) -2\psi(1+u+v))^2 \right]   \G^2(-u)\G^2(-v)\G^2(1+u+v) = \\
\oint_C du~dv~x^u~y^v \left[\le \ln x -2\psi(-u) + 2\psi(1+u+v)\ri \le \psi(-u) - \psi(1+u+v)\ri +\right.\\
\left. + \le \ln y -2\psi(-v) + 2\psi(1+u+v)\ri \le \psi(-v) - \psi(1+u+v)\ri \right.\\
\left. + (\psi(-u) + \psi(-v) -2\psi(1+u+v))^2 \right]   \G^2(-u)\G^2(-v)\G^2(1+u+v) = \\
\oint_C du~dv~x^u~y^v \left[\ln x \le \psi(-u) - \psi(1+u+v)\ri + \ln y \le \psi(-v) - \psi(1+u+v)\ri \right.\\
\left. - (\psi(-u)-\psi(-v))^2\right] \G^2(-u)\G^2(-v)\G^2(1+u+v) = \\
\frac{1}{2}(\ln^2 x +\ln^2 y)\oint_C du~dv~x^u~y^v \G^2(-u)\G^2(-v)\G^2(1+u+v) \\
- \oint_C du~dv~x^u~y^v(\psi(-u)-\psi(-v))^2 \G^2(-u)\G^2(-v)\G^2(1+u+v).
\end{eqnarray*}
The first term is proportional to the first UD function, and the result for integral (\ref{not implicit-1}) is 
\begin{eqnarray}
\lim_{\ve_2 \rar 0, \ve_1 \rar 0} \oint_C du~dv~x^u~y^v M_2^{(u,v)}(\ve_1,\ve_2,\ve_3) = \no\\
\le\frac{1}{4}\ln^2 \frac{x}{y}-  \frac{3}{2}\le\G(1-\ve)\G(1+ \ve) \ri^{(2)}_{\ve=0}\ri 
\oint_C du~dv~x^u~y^v \G^2(-u)\G^2(-v)\G^2(1+u+v) - \no\\
- \frac{3}{2}\oint_C du~dv~x^u~y^v  \le \psi(-u) - \psi(-v)\ri^2 \G^2(-u)\G^2(-v)\G^2(1+u+v). \label{ssylka-3} 
\end{eqnarray}
This result is related to formula (\ref{nah}) via integration by parts in the complex planes of $u$ and $v$.

\subsection{Explicit two-fold MB transform for the three-rung ladder}

We need to derive the same limit for  $M_3^{(u,v)}(\ve_1,\ve_2,\ve_3).$  We show that the result is reduced to 
the Euler $\psi$ function and its derivatives. According to Eq. (\ref{implicit-2}),  
\begin{eqnarray}
\lim_{\ve_2 \rar 0, \ve_1 \rar 0} M_3^{(u,v)}(\ve_1,\ve_2,\ve_3) = \no\\
\lim_{\ve_2 \rar 0, \ve_1 \rar 0} \left[\frac{1}{\ve_1\ve_2} M_2^{(u+\ve_1,v+\ve_2)}(\ve_1,\ve_2,\ve_3) + 
\frac{J}{\ve_2\ve_3} M_2^{(u+\ve_1,v)}(\ve_1) +  
\frac{J}{\ve_1\ve_3} M_2^{(u,v+\ve_2)}(\ve_2) \right] = \no\\
\no\\
\lim_{\ve_2 \rar 0, \ve_1 \rar 0} \left[\frac{1}{\ve_1\ve_2} M_2^{(u+\ve_1,v+\ve_2)}(\ve_1,\ve_2,\ve_3) -
\frac{J}{\ve_2(\ve_1+\ve_2)} M_2^{(u+\ve_1,v)}(\ve_1) \right. \no\\
\left. 
- \frac{J}{\ve_1(\ve_1+\ve_2)} M_2^{(u,v+\ve_2)}(\ve_2) \right]= \no\\
\lim_{\ve_2 \rar 0, \ve_1 \rar 0} \left(\frac{1}{\ve_1\ve_2} \left[M_2^{(u+\ve_1,v+\ve_2)}(\ve_1,\ve_2,\ve_3) -
J M_2^{(u+\ve_1,v)}(\ve_1) \right] 
+  \frac{1}{\ve_1^2}\left[M_2^{(u+\ve_1,v)}(\ve_1) - M_2^{(u,v)} \right] \right). 
\label{implicit-4}
\end{eqnarray}
We use the notation   
\begin{eqnarray*}
\le \G(-\ve_i-u)\G(\ve_i-v)\ri^{(2)}_0,~~~ 
\le \G(-\ve_i-u)\G(\ve_i-v)\ri^{(3)}_0, ~~~
\le \G(-\ve_i-u)\G(\ve_i-v)\ri^{(4)}_0 
\end{eqnarray*}
for the corresponding derivatives with respect to $\ve_i$ at the point $\ve_i= 0.$  
We obtain for the second term of (\ref{implicit-4}) the expansion in terms of $\ve_1$
\begin{eqnarray*}
M_2^{(u+\ve_1,v)}(\ve_1) - M_2^{(u,v)} = \\
\G^2(1+u+v)\left[ \frac{\ve_1}{6}\le \G(-\ve_1-u)\G(\ve_1-v)\ri^{(3)}_0 \G(-u)\G(-v) \right.\\
\left. - \frac{\ve_1}{2} \le \G(-\ve_1-u)\G(\ve_1-v) \ri'_0 \le \G(-\ve_1-u)\G(\ve_1-v)\ri^{(2)}_0 \right.\\
\left.
+ \ve_1 a_0^{(2)} \le\G(-\ve_1-u)\G(\ve_1-v)\ri'_0 \G(-u)\G(-v) 
+ \frac{\ve_1^2}{24}\le \G(-\ve_1-u)\G(\ve_1-v)\ri^{(4)}_0 \G(-u)\G(-v)  \right.\\
\left.
- \frac{\ve_1^2}{6} \le \G(-\ve_1-u)\G(\ve_1-v) \ri^{(3)}_0 \le \G(-\ve_1-u)\G(\ve_1-v)\ri'_0 \right.\\
\left.
+ a_0^{(2)} \ve_1^2\Bigl(\frac{3}{4}\le\G(-\ve_1-u)\G(\ve_1-v)\ri^{(2)}_0 \G(-u)\G(-v) \Bigr.\right.\\
\Bigl.\left.
- \frac{1}{2}\le \G(-\ve_1-u)\G(\ve_1-v) \ri'_0 \le \G(-\ve_1-u)\G(\ve_1-v) \ri'_0 \Bigr) 
+ a_0^{(4)} \ve_1^2\frac{5}{24}\G^2(-u)\G^2(-v)\right] + o(\ve_1^2). 
\end{eqnarray*}

We use the notation 
\begin{eqnarray*}
a(\ve) = \left[ \G(1-\ve)\G(1+\ve)\right]^{-1},  ~~ a_0^{(n)} = (a(\ve))^{(n)}_{\ve=0}.
\end{eqnarray*}

Going back to the first term of Eq. (\ref{implicit-4}) we derive, by using Eq. (\ref{ssylka}), that
\begin{eqnarray}
M_2^{(u+\ve_1,v+\ve_2)}(\ve_1,\ve_2,\ve_3)  = \no\\
 J\left[\frac{1}{\ve_1\ve_2}\frac{\G(-\ve_1-u)\G(-\ve_2-v)\G(\ve_2 -u)\G(\ve_1  -v)\G^2(1 + u + v)}
{\G(1-\ve_3)\G(1+ \ve_3)} \right.\no\\
- \left.\frac{1}{\ve_2(\ve_1+\ve_2)}  \frac{\G(-\ve_1-u)\G(-v)\G(-u)\G(\ve_1 - v)\G^2(1 + u + v)}
{\G(1-\ve_1)\G(1+ \ve_1)} \right. \no\\
- \left.\frac{1}{\ve_1(\ve_1+\ve_2)}\frac{\G(-u)\G(-\ve_2-v)\G(\ve_2-u)\G( -v)\G^2(1 + u + v)}{\G(1-\ve_2)\G(1+ \ve_2)} \right].
\label{explicit-2}
\end{eqnarray}

A simple calculus gives that 
\begin{eqnarray*}
M_2^{(u+\ve_1,v+\ve_2)}(\ve_1,\ve_2,\ve_3) - JM_2^{(u+\ve_1,v)}(\ve_1) =  \\
\ve_2 \G^2(1+u+v)\left[ - \frac{1}{6}\le \G(-\ve_1-u)\G(\ve_1-v)\ri^{(3)}_0 \G(-u)\G(-v) \right.\\
\left. + \frac{1}{2} \le \G(-\ve_1-u)\G(\ve_1-v) \ri'_0 \le \G(\ve_2-u)\G(-\ve_2-v)\ri^{(2)}_0 \right.\\
\left.
- a_0^{(2)} \le\G(-\ve_1-u)\G(\ve_1-v)\ri'_0 \G(-u)\G(-v) \right.\\
\left. - \frac{\ve_1}{24}\le \G(-\ve_1-u)\G(\ve_1-v)\ri^{(4)}_0 \G(-u)\G(-v)  \right.\\
\left.
+ \frac{\ve_1}{4} \le \G(-\ve_1-u)\G(\ve_1-v) \ri^{(2)}_0 \le \G(\ve_2-u)\G(-\ve_2-v)\ri^{(2)}_0 \right.\\
\left.
+ a_0^{(2)} \ve_1\Bigl(\frac{1}{4}\le\G(-\ve_1-u)\G(\ve_1-v)\ri^{(2)}_0 \G(-u)\G(-v) \Bigr.\right.\\
\Bigl.\left.
- \le \G(-\ve_1-u)\G(\ve_1-v) \ri'_0 \le \G(-\ve_1-u)\G(\ve_1-v) \ri'_0 \Bigr) 
+ a_0^{(4)} \ve_1\frac{5}{24}\G^2(-u)\G^2(-v) + o(\ve_1)\right] + o(\ve_2). 
\end{eqnarray*}

Thus, we obtain the limit for  $M_3^{(u,v)}(\ve_1,\ve_2,\ve_3)$ when the $\ve$-terms are vanishing 
\begin{eqnarray}
\lim_{\ve_2 \rar 0, \ve_1 \rar 0}  M_3^{(u,v)}(\ve_1,\ve_2,\ve_3) = \no\\
\lim_{\ve_2 \rar 0, \ve_1 \rar 0} \le\frac{1}{\ve_1\ve_2} \left[M_2^{(u+\ve_1,v+\ve_2)}(\ve_1,\ve_2,\ve_3) -
JM_2^{(u+\ve_1,v)}(\ve_1) \right] \right.\no\\
\left.
+  \frac{1}{\ve_1^2}\left[M_2^{(u+\ve_1,v)}(\ve_1) - M_2^{(u,v)} \right]\ri = \no\\
\G^2(1 + u + v)
\left[  \frac{1}{4}\le \G(-\ve_1-u)\G(\ve_1-v)\ri^{(2)}_0 \le \G(-\ve_1-u)\G(\ve_1-v)\ri^{(2)}_0  \right.\no\\
\left.
- \frac{1}{6}\le \G(-\ve_1-u)\G(\ve_1-v)\ri^{(3)}_0 \le \G(-\ve_1-u)\G(\ve_1-v)\ri'_0 \right.\no\\
\left.
+ a_0^{(2)} \Bigl(-\frac{3}{2}\left(\le\G(-\ve_1-u)\G(\ve_1-v)\ri'_0\right)^2 - \G(-u)\G(-v) \le \G(-\ve_1-u)\G(\ve_1-v) \ri^{(2)}_0 \Bigr) \right.\no\\
\left.
+ a_0^{(4)} \frac{5}{12}\G^2(-u)\G^2(-v)\right].  \label{ssylka-4}
\end{eqnarray}

\section{Conclusions}

In this paper we have explicitly calculated a new type of the two-fold MB integrals. 
The integrals of that type have been obtained due to recursive relations of the momentum loop integrals
which are derived  from the Belokurov-Usyukina loop reduction method. The recursion creates an infinite system of the integral relations, which allowed us to 
represent the MB transform of momentum $(L+1)$-loop integral as a linear combination of the MB transforms of momentum $L$-loop integrals. All the integrals have been 
reduced to the Appell's hypergeometric function corresponding to the one-loop momentum massless integral. 
In order to derive those relations, we needed to shift the complex variables 
of the MB transforms on the right hand sides of the integral relations. This shift is necessary 
in order to organize the same dependence on the parameters $x$ and $y$ in the integrands on the  
on the left and on the right hand sides. As the result, in the integral relations
representation (\ref{nah}) appears while representation (\ref{ssylka-3}) does not appear.

At the first site, the results (\ref{ssylka-3}) and (\ref{ssylka-4}) are difficult to analyse, since on the 
right hand sides of them certain products of higher order derivatives of the Euler gamma functions 
\begin{eqnarray}
\frac{d^n}{d \ve^n} \G(-\ve-u)\G(\ve-v)|_{\ve=0}  \label{der-1}
\end{eqnarray}
appear. These derivatives can be re-written as 
\begin{eqnarray}
\frac{d^n}{d\ve^n} \G(-\ve-u)\G(\ve-v)|_{\ve=0} = (\pd_u-\pd_v)^n  \G(-u)\G(-v),  \label{der-2}
 \end{eqnarray}
that makes them more convenient for use in the integration by parts.

However, the integration by parts in the planes of complex variables $u$ and $v$ can serve to us 
just to prove the coincidence of two, at first view, different representations. The reason why 
the representation for MB transforms of UD functions should be a simple expansion in terms of 
derivatives  (\ref{der-1}) or (\ref{der-2}) is in a recursive construction of the MB images. 
The recursive construction has nothing to do with quantum field theory or with theory of polylogarithms. 
We will prove this observation further. For this purpose representation (\ref{ssylka-3}) is more useful than 
representation (\ref{nah}) used in the integral relations \footnote{ Actually, beyond 
the four rung in the ladder the difference between two types of representation disappears}.

As the first step, we consider the integral relation \footnote{This relation can easily be derived via shifts in the planes of complex integration variables $u$ and $v$ 
in the MB transforms} that can be found in Ref. \cite{Usyukina:1992jd}
\begin{eqnarray*}
J(1,1,1-\ve) = \frac{1}{2}(p_3^2)^{\ve}\left[x^{\ve}J(1-\ve,1+\ve,1) + y^{\ve}J(1+\ve,1-\ve,1) \right], 
\end{eqnarray*}
or, equivalently,
\begin{eqnarray}
\oint_C du~dv~x^u~y^v D^{(u,v)}[1,1,1-\ve] = \no\\
\frac{1}{2} \oint_C du~dv~x^u~y^v\Bigl(x^{\ve}D^{(u,v)}[1-\ve,1+\ve,1] + y^{\ve}D^{(u,v)}[1+\ve,1-\ve,1] \Bigr). \label{structure} 
\end{eqnarray}

The benefits of representation (\ref{structure}) are in more simple dependence on the variable $\ve,$
namely, it presents in the numerator in two factors only, 
\begin{eqnarray*}
D^{(u,v)}[1-\ve,1+\ve,1] = \frac{\G(-u)\G(-v) \G(-\ve-u)\G(\ve - v) \G^2(1+u+v)}{\G(1-\ve)\G(1+\ve)}.
\end{eqnarray*}
For example, the equation for $M_2$ can be re-written as 
\begin{eqnarray*}
\lim_{\ve_2 \rar 0, \ve_1 \rar 0} \oint_C du~dv~x^u~y^v M_2^{(u,v)}(\ve_1,\ve_2,\ve_3) = \no\\
\lim_{\ve_2 \rar 0, \ve_1 \rar 0} \oint_C du~dv~x^u~y^v \Bigl[\frac{1}{\ve_2\ve_3} y^{\ve_2} D^{(u,v)}[1-\ve_1] + \frac{1}{\ve_1\ve_2} D^{(u,v)}[1+\ve_3] 
+ \frac{1}{\ve_1\ve_3} x^{\ve_1} D^{(u,v)}[1-\ve_2] \Bigr] = \no\\
\lim_{\ve_2 \rar 0, \ve_1 \rar 0} \oint_C du~dv~x^u~y^v \left[ \frac{a(\ve_1)}{\ve_2\ve_3} y^{\ve_2} \left[ x^{\ve_1}\Gamma(-u-\ve_1)\Gamma(-v+\ve_1) 
+ y^{\ve_1}\Gamma(-u+\ve_1)\Gamma(-v-\ve_1) \right]  \right. \no\\
\left. 
+ \frac{a(\ve_3)}{\ve_1\ve_2} \left[ x^{-\ve_3}\Gamma(-u+\ve_3)\Gamma(-v-\ve_3) 
+ y^{-\ve_3}\Gamma(-u-\ve_3)\Gamma(-v+\ve_3) \right] \right. \no\\
\left. 
+ \frac{a(\ve_2)}{\ve_1\ve_3} x^{\ve_1} \left[ x^{\ve_2}\Gamma(-u-\ve_2)\Gamma(-v+\ve_2) 
+ y^{\ve_2}\Gamma(-u+\ve_2)\Gamma(-v-\ve_2) \right]\right] \times \no\\
\times\G(-u)\G(-v) \G^2(1+u+v). 
\end{eqnarray*}
In the limit of vanishing $\ve_i$ the value of this expression contains derivatives (\ref{der-1}) only, that is,
\begin{eqnarray*}
\lim_{\ve_2 \rar 0, \ve_1 \rar 0} \oint_C du~dv~x^u~y^v M_2^{(u,v)}(\ve_1,\ve_2,\ve_3) = \\
\oint_C du~dv~x^u~y^v \G(-u)\G(-v) \G^2(1+u+v)\left[\frac{3}{2}\le a(\ve)\G(-u-\ve)\G(-v + \ve)\ri^{(2)}_0 + \right. \\
\left. \frac{3}{2}\ln\frac{x}{y}\le a(\ve)\G(-u-\ve)\G(-v + \ve)\ri'_0 + \frac{1}{4}\ln^2\frac{x}{y}\G(-u)\G(-v)\right].
\end{eqnarray*}
This expression does not have a form of (\ref{nah}) or (\ref{ssylka-3}), however it can be transformed to 
that form by using the integration by parts in the complex planes of $u$ and $v.$ 

In  integral relation (\ref{implicit-2}) for  $M_3,$ which is 
\begin{eqnarray*}
M_3^{(u,v)}(\ve_1,\ve_2,\ve_3) = \no\\
\frac{1}{\ve_1\ve_2} x^{-\ve_1} y^{-\ve_2} M_2^{(u,v)}(\ve_1,\ve_2,\ve_3) + \frac{J}{\ve_2\ve_3} x^{-\ve_1} M_2^{(u,v)}(\ve_1) 
+  \frac{J}{\ve_1\ve_3} y^{-\ve_2} M_2^{(u,v)}(\ve_2), 
\end{eqnarray*}
we substitute the structure found in (\ref{structure}), and again in the limit of vanishing $\ve_i$ 
we obtain an expansion in terms of derivatives (\ref{der-1}), that is,
\begin{eqnarray}
\lim_{\ve_2 \rar 0, \ve_1 \rar 0}  \oint_C du~dv~x^u~y^v M_3^{(u,v)}(\ve_1,\ve_2,\ve_3) = \no\\
\oint_C du~dv~x^u~y^v \G(-u)\G(-v) \G^2(1+u+v)\left[\frac{5}{12} \le a(\ve)\G(-u-\ve)\G(-v + \ve)\ri^{(4)}_0 + \right. \no\\
\left.
\frac{5}{6}\ln \frac{x}{y} \le a(\ve)\G(-u-\ve)\G(-v + \ve)\ri^{(3)}_0 + 
\frac{1}{2}\ln^2 \frac{x}{y} \le a(\ve)\G(-u-\ve)\G(-v + \ve) \ri^{(2)}_0 \right. \no\\
\left.
+ \frac{1}{12}\ln^3 \frac{x}{y} \le a(\ve)\G(-u-\ve)\G(-v + \ve)\ri'_0  \right]. \label{expansion}
\end{eqnarray}
By using the integration by part in the complex plane of $u$ and $v$, we obtain  coincidence 
with Eq. (\ref{ssylka-4}). 

Following the recursive procedure for higher values of $n,$ we will obtain the 
higher derivatives of type (\ref{der-1}) multiplied by powers of $\ln  \frac{x}{y}.$ 
At present, it is difficult to calculate the coefficients in front of the expansion terms. 
However, it is clear that the finiteness of the limit of vanishing $\ve_i$ has nothing to do with 
the polylogarithms, since instead of the Euler $\G$ function we can write any other smooth function
in construction (\ref{structure}) and the limit remains finite. Thus, the infinite sum in the limit of vanishing $\ve_i$ of 
the quantities  $M_n^{(u,v)}(\ve_1,\ve_2,\ve_3)$ constructed from  (\ref{structure}) has underlying integrable 
structure which can be uncovered by identifying coefficients in front of the terms in the expansion 
(\ref{expansion}) for higher $n.$ That integrable structure has nothing common with the MB transforms of
polylogarithms and is based on the properties that can be studied by the basic methods of mathematical 
analysis.

\subsection*{Acknowledgments}

I.K. was supported by Fondecyt (Chile) grants 1040368, 1050512 and
by DIUBB grant (UBB, Chile) 102609. E.A.N.C. was supported by
Direcci\'on de Investigaci\'on de la Universidad de La Serena, DIULS
CD11103. M.R.M. is supported by project Nro. MTM2009-12927,
Ministerio de Ciencia e Innovaci\'on, Espa\~na and by project Nro. 1080628,
Fondecyt (Chile). P.A. thanks Physics Department of Facultad de
Ciencias Fisicas y Matematicas, UdeC, Chile for financial support of his
travels from Concepcion to Chill\'an. This paper is based on the lectures ``Integraci\'on multiple
y sus aplicaciones'' given by I.K. at Mathematical department of Facultad de
Educaci\'on y Humanidades, UBB, Chillan from September 2010 till January 2011.
He is grateful to all the students attended to that course for their
questions and interest.  A part of the results was presented at XXV Jornada 
de Matem\'atica de la Zona Sur, Concepci\'on, Chile, April 2012.

\end{document}